\documentclass[sigconf]{acmart}

\usepackage{graphicx}

\usepackage{amssymb}
\usepackage{pifont}
\usepackage{booktabs}
\usepackage{multirow}
\usepackage{url}
\AtBeginDocument{%
  }

\newcommand{\q}[1]{\textit{``#1''}}

\newcommand{\tool}{Spatula}

\newcommand{\revise}[1]{#1}

\usepackage{enumitem}
\copyrightyear{2026}
\acmYear{2026}
\setcopyright{cc}
\setcctype{by}
\acmConference[UIST '26]{The 39th Annual ACM Symposium on User Interface Software and Technology}{November 02--05, 2026}{Detroit, MI, USA}
\acmBooktitle{The 39th Annual ACM Symposium on User Interface Software and Technology (UIST '26), November 02--05, 2026, Detroit, MI, USA}
\acmDOI{10.1145/3830398.3830633}
\acmISBN{979-8-4007-2856-3/2026/11}

\begin{document}


\title{\tool: Exploring On-Demand In-Situ Interfaces and Interaction for Attribute Control}


\author{Boyu Li}
\affiliation{%
  \institution{
  The Hong Kong University of Science and Technology}
  \city{Hong Kong}
  \country{China}}
\email{blibr@connect.ust.hk}

\author{Linjie Qiu}
\affiliation{%
  \institution{The Hong Kong University of Science and Technology (Guangzhou)}
  \city{Guangzhou}
  \country{China}}
\email{lqiu250@connect.hkust-gz.edu.cn}

\author{Lin-Ping Yuan}
\affiliation{%
  \institution{The Hong Kong University of Science and Technology}
  \city{Hong Kong}
  \country{China}}
\email{yuanlp@cse.ust.hk}

\author{Duotun Wang}
\affiliation{%
  \institution{The Hong Kong University of Science and Technology (Guangzhou)}
  \city{Guangzhou}
  \country{China}}
\email{dwang866@connect.hkust-gz.edu.cn}

\author{Yue Jiang}
\affiliation{%
  \institution{University of Utah}
  \city{ Salt Lake City}
  \country{USA}}
\email{yue.jiang@utah.edu}

\author{Zeyu Wang}
\affiliation{%
  \institution{The Hong Kong University of Science and Technology (Guangzhou)}
  \city{Guangzhou}
  \country{China}}
\affiliation{%
  \institution{The Hong Kong University of Science and Technology}
  \city{Hong Kong}
  \country{China}}
\email{zeyuwang@ust.hk}

\author{Hongbo Fu}
\authornote{Corresponding author.}
\affiliation{%
  \institution{
  The Hong Kong University of Science and Technology}
  \city{Hong Kong}
  \country{China}}
\email{fuplus@gmail.com}



\begin{abstract}
Controlling attributes is a critical step toward achieving the final creative outcome, yet current approaches fall short in supporting users in the iterative refinement of generative content. We propose Spatula, a proof-of-concept system that generates on-demand, in-situ attribute control interfaces and interactions for creating motion graphics. Building on a technical probe that automatically analyzes animation context and generates corresponding attributes and UI, we frame attribute control as an explorable landscape and explore the attribute control space along four key dimensions: Discoverability, Resolution, Scope, and Expandability. Findings from a user study (N=12) show that our system provides intuitive and convenient interactions while supporting diverse needs for fine-grained parameter control. Furthermore, our applications demonstrate that the plug-and-play design generalizes to other domains, such as web design and 3D modeling.
\end{abstract}



\begin{teaserfigure}
  \centering
  \includegraphics[width=1\textwidth]{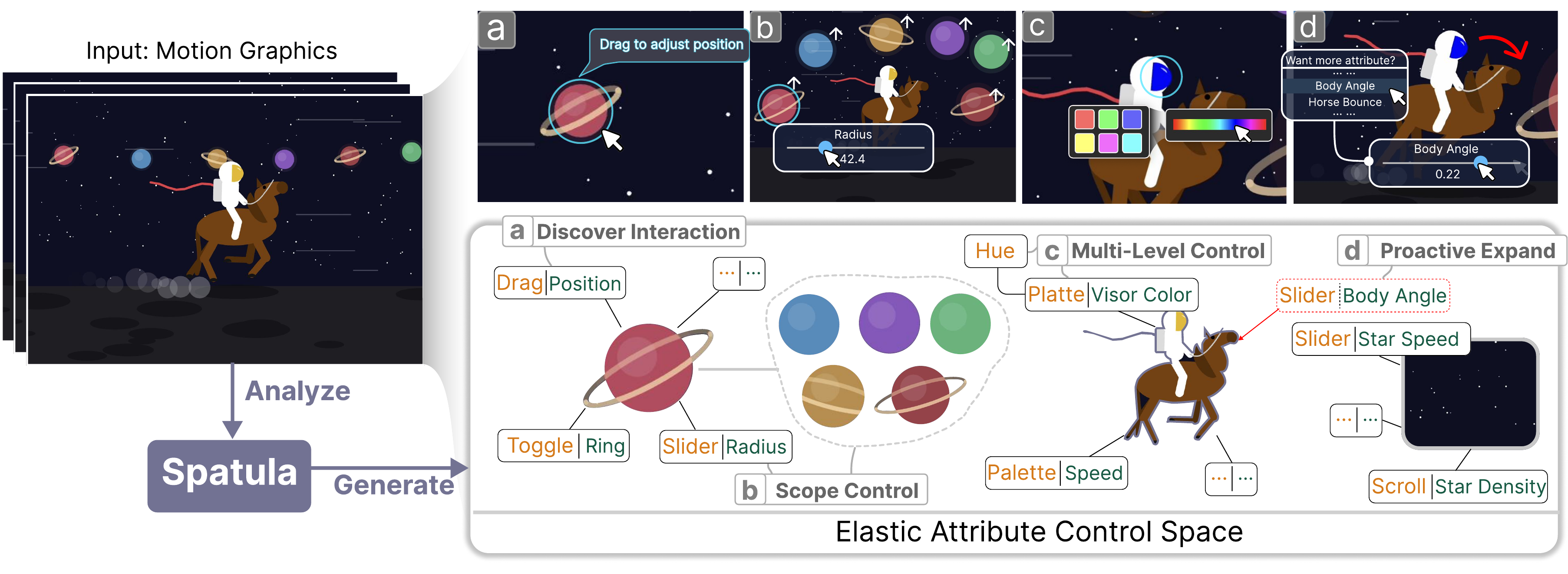}
  \caption{We introduce \tool, a system for generating on-demand, in-situ attribute control interfaces for motion graphics. Given a motion graphic, \tool~ constructs an interactive attribute control scaffold that operationalizes an Elastic Attribute Control Space along four dimensions : (a) discoverability via context-aware in-situ hints, (b) scope control for semantically coordinated manipulation, (c) multi-resolution adjustment for varying precision, and (d) proactive attribute space expansion.}
  \Description{}
  \label{fig:teaser}
\end{teaserfigure}



\maketitle
\section{Introduction}
While generative AI has significantly lowered the barrier to content creation~\cite{liu2024how}, refining generated results remains a major challenge. Achieving a specific vision often requires precise, parameter-level adjustments to attributes such as color, motion, or scale. In current LLM-based workflows, these refinements rely heavily on text prompts~\cite{masson2024directgpt}, which are indirect, lack immediate feedback loops, and provide poor predictability. For example, prompting an animation to \q{move faster} offers no guarantee of the magnitude of change or which specific element will be modified. As a result, fine-grained control in generative workflows remains disconnected from the visual canvas.

The challenge of attribute control, however, is not new. Traditional authoring tools expose rich parameter spaces through hierarchical panels, enabling precise manipulation but at the cost of high cognitive load~\cite{oviatt2006human}. In contrast, alternative paradigms such as sketch-based~\cite{shen2024neural,suzuki2020realitysketch} or tangible interfaces~\cite{li2024anicraft,held20123d} simplify interaction by abstracting parameters behind intuitive metaphors, yet often sacrifice the precision required for detailed adjustments. Across these paradigms, a persistent tension remains between usability and control.

Recent advances in LLM-based program generation have introduced on-demand interfaces, where controls are dynamically generated based on context and user intent~\cite{zhu2025agentAR,ye2026mographGPT}. Prior work has explored this paradigm across domains such as education~\cite{augmented2024gunturu}, AR/VR~\cite{zhu2025agentAR}, and collaborative systems~\cite{xia2023crosstalk}, with emerging applications in creative authoring. However, existing authoring systems treat on-demand control as an external and ad-hoc addition rather than an integrated part of the creative process~\cite{xie2025datawink,ye2026mographGPT}. First, generated widgets are often placed in disconnected side panels, separated from the visual context and lacking support for in-situ interaction and direct manipulation~\cite{Ben1981directmanipulation}. Second, widget generation is typically outsourced to black-box LLM analysis that fails to consider what controls are needed to support user intent or how they should be organized to enable effective refinement.

We argue that the core challenge is not how to generate interfaces, but how to structure control to make it more actionable and intuitive. Instead of viewing controls as ad-hoc UI elements, we frame attribute control as an explorable and dynamically constructed Elastic Attribute Control Space. In this space, relationships between parameters, interaction mappings, and levels of granularity are explicitly organized. This perspective shifts the role of LLMs from producing isolated interface elements to constructing interactive control scaffolds that support iterative refinement. 

To ground this concept, we focus on motion graphics creation, as they are widely explored in LLM-based generation~\cite{liu2025logomotion}, involve spatiotemporal transformations common across many creative tasks~\cite{ma2025mover}, and expose explicit and interpretable attributes that support parameterized control~\cite{shi2024piet}, making them well-suited for studying fine-grained control in generative workflows. Enabling effective refinement in this context requires resolving how user actions map to underlying parameters across different scopes and levels of control, giving rise to an attribute control space. While such a space is central to iterative editing, how to dynamically construct and interact with it in practice remains underexplored.

We conducted two formative studies to explore this problem. First, we interviewed professional animators to compare LLM-based and traditional workflows, identifying key needs in iterative refinement. Next, we developed a tech probe that automatically generate control UI to motion graphics for in-situ editing. Findings from this probe reveal structural gaps in how attribute control is exposed, motivating a design that adapts to user intent and visual context while maintaining interface simplicity.
Building on these insights, we introduce \tool, an in-situ authoring interface that operationalizes the Elastic Attribute Control Space.
\revise{To reliably translate animation logic into runtime attribute-control mappings, \tool~ uses an external skill base derived from commercial authoring software.} Rather than relying on a fixed control space, \tool\ lets users explore and reshape attributes control through adaptive interaction strategies. It provides context-aware hints (Discoverability), multi-resolution controls (Resolution), collective adjustment across related elements (Scope), and proactive attribute suggestions (Expandability), making the control space both actionable and flexible.

Through a comparative user study (N=12), we show that \tool supports users in exploring and utilizing attribute control more effectively than baseline methods. The system enables low-latency, fine-grained manipulation while maintaining a lightweight interface, supporting both deliberate refinement and open-ended exploration. \revise{Moreover, although our prototype focuses on code-based motion graphics, we demonstrate its applicability to other program-based visual creation contexts, such as web design and 3D modeling.}
We make the following contributions:
\begin{itemize}
    \item \revise{We identify the lack of structured control as a key limitation in on-demand generated attribute control and reframe it as an explorable and structured space.}
    \item We introduce the concept of an elastic attribute control space and derive mechanisms (Discoverability, Resolution, Scope, and Expandability) for constructing and interacting.
    \item We present \tool, an on-demand, in-situ interface that enables real-time, fine-grained parameter manipulation via context-aware control scaffolds.
    \item We provide empirical insights from a user study into how users explore and utilize attribute control across different levels of expertise. 

\end{itemize}

\section{Related Work}

\subsection{On-demand Interface}

The design philosophy of on-demand interfaces originates from the balance between system expressivity and cognitive bandwidth of users. In complex authoring environments, exposing all available controls simultaneously leads to UI clutter, which increases visual search time and cognitive overhead~\cite{mcgrenere2002thesis, shneiderman2016dui}. To address this, the HCI community has explored strategies to reveal interface elements only when they are relevant to the user’s immediate goals.

Early explorations pioneered on-demand interfaces by using sensor-mediated touch detection to reveal toolbars or menus in response to a user’s hand proximity or context, reducing the need for explicit command invocation~\cite{ken1999touchsensing}. FlowMenu~\cite{francois2000FlowMenu} and Marking Menus~\cite{zhao2004markingmenus} translate command selection into fluid gesture, effectively reducing the temporal gap between intent and execution.

In the realm of user interfaces, progressive disclosure has become a standard design pattern to manage complexity by deferring advanced features to secondary layers~\cite{tidwell2010designing}. However, traditional progressive disclosure often relies on static hierarchies (e.g., nested menus), which may not adapt to dynamic workflows. Research in adaptive and predictive UIs leverages user behavior logs or machine learning to anticipate which parameters are most likely to be tuned~\cite{gajos2010supple,song2025preguidedMOUI}. Recent work integrates user conversational context with LLMs to dynamically generate on-demand interfaces, providing a more adaptive alternative to traditional optimization framework~\cite{chen2025generativeui}. Datawink also implements a conversational agent to interpret adaptation goals and generate UI widgets such as sliders and color pickers, translating high-level textual prompts into structured and on-demand UI components~\cite{xie2025datawink}. 
\revise{Related work has also explored dynamically constructed and malleable controls. DynaVis synthesizes persistent widgets from natural language visualization edits~\cite{vaithilingam2024dynavis}, while malleable overview-detail interfaces allow users to surface, transform, and generate fluid attributes across views~\cite{min2025malleable}. Spatula extends this direction to executable animations, focusing on how generated controls are exposed and organized in situ.}

While existing on-demand interfaces excel at command invocation, they often fail to provide the visual scaffolding needed to reason about high-dimensional, interdependent parameters. Our work, Spatula, builds upon the in-situ interaction paradigm~\cite{bubblingmenus2007tsandilas} but specifically focuses on the on-demand generation of semantic interface overlays. By surfacing controls directly over the canvas, we aim to bridge the gap between abstract parameter spaces and concrete visual outcomes. Our approach operationalizes the Elastic Attribute Control Space by transforming latent parameters into actionable, in-situ scaffolds that adapt to user's immediate intent.


\subsection{Parameter Tuning in Authoring}

Parameter tuning poses a significant challenge in authoring systems where users must reason about complex mental models. Traditionally, users relied on iterative trial-and-error with direct manipulation tools like sliders and knobs~\cite{donovan2015designscape, michael2012designmethod}. Early work on direct manipulation emphasized immediate, reversible actions and continuous feedback as key principles for effective parameter control~\cite{Ben1981directmanipulation}. Sliders, knobs, and visual controls became standard mechanisms for exposing parameters, particularly in domains such as graphics and animation~\cite{brad1998hcihistory}. However, as systems grew more expressive, the number and interdependence of parameters increased, making exhaustive manual tuning impractical. 

To mitigate this, industry software has adopted higher-level abstractions that map raw parameters to spatial or semantic controls. For example, color curves in Adobe Photoshop condense pixel-level manipulations into geometric interactions~\cite{adobeRGB2026}, Macro Controls in digital audio workstations aggregate interdependent parameters into single, goal-oriented knobs~\cite{macroControlsDAW2026}, and interactive color wheels in design software provide an intuitive circular interface for selecting and combining colors from underlying RGB values~\cite{figmaColorWheel2026}.

Prior work has also explored attribute-centered direct manipulation. Object-Oriented Drawing reifies visual attributes as manipulable objects that can be moved, copied, and associated with drawing elements~\cite{xia2016object}. Collection Objects extends this approach to fluid formation and manipulation of aggregate selections~\cite{xia2017collection}, while DataInk supports direct binding between visual and data attributes through pen-and-touch interaction~\cite{xia2018dataink}. These systems establish important techniques for directly manipulating attributes and groups.

Beyond static abstractions, HCI research has explored algorithmic approaches to further reduce the cost of parameter search. Interactive optimization techniques~\cite{yuki2022Bo, liao2025conhitpop} model user preference as a black-box function, allowing users to converge toward desirable configurations through lightweight comparisons rather than explicit parameter adjustments. It has been applied to wide tasks, including visual~\cite{yuki2016Selph} and interaction~\cite{niwa2025codesignop} design. Recently, the integration of Large Language Models (LLMs) has further transformed this landscape by shifting from heuristic-based search to intent-driven program synthesis~\cite{chen2021codex, austin2021program}. Systems like VLMaterial~\cite{li2025vlmaterial}, UICoder~\cite{wu2024uicoder}, and Athena~\cite{beason2025athena} demonstrate how LLMs can directly translate high-level natural language into structured interface code or procedural material graphs.
However, a critical limitation remains across these advancements: both traditional algorithmic optimization and current LLM-based synthesis systems essentially act as black boxes. They typically treat synthesis as a construction task, generating a final webpage or chart for consumption rather than for manipulation. While this intent-driven mapping is highly expressive, it obscures the relationship between user input and the resulting low-level configurations. Consequently, users often struggle to develop accurate mental models of the design space and perform the fine-grained, localized adjustments necessary for continuous exploration and refinement.
\revise{Recent systems such as Narrative Motion Blocks and SketchDynamics use geometric input and direct manipulation to express animation intent, but focus on motion creation rather than the real-time exploration and tuning of underlying parameters~\cite{bourgault2025narrative}.}

Unlike these systems, Spatula seeks to combine the strengths of both traditional GUI-based parameter exposure and implicit, algorithmic tuning. By generating on-demand, in-situ parameter control widgets, Spatula avoids the cognitive load associated with massive, isolated control panels while still allowing users to directly inspect and adjust relevant parameters in real time.

\section{Formative Study}
We conducted a two-stage formative study to ground our design in practical motion graphics workflows. In Stage 1, we combined semi-structured interviews with design elicitation tasks to identify bottlenecks in both manual and AI-assisted creation. We focused on the post-generation refinement phase, where designers adjust element-level attributes to achieve precise intent.
These findings informed Stage 2: the development of a technical probe. We implemented an LLM-driven system that generates on-demand, in-situ control interfaces for motion graphics. Deploying this probe allowed us to observe real-time interactions and derive more concrete design requirements for the final system

\subsection{Problem Understanding in Current Practices}
With an initial design idea in mind, the first stage served as an exploratory inquiry to better understand designers’ real needs in practice.

\subsubsection{Participants and Procedure.}
Following prior work~\cite{videocraft2025li,augmented2024gunturu}, we recruited six participants (P1–P6; two male, four female) from the local university community (mean age = 27.1, SD = 3.4). All were right-handed and had 1–3 years of experience creating motion graphics with tools such as After Effects~\cite{ae} and Figma~\cite{figmaColorWheel2026}. We intentionally recruited experienced participants to enable informed reflections on both conventional workflows and emerging AI-assisted practices. Some had prior exposure to LLM-based approaches, including generating p5.js animations (P1, P5) and using AI tools such as Fogsight (P2).
The study was conducted offline in two phases. First, semi-structured interviews examined attribute adjustment in traditional softwares, focusing on iterative parameter refinement, strategies, and challenges. Second, we conducted a design elicitation study on LLM-based workflows: participants generated p5.js animations using a text-based LLM (Gemini 3), iteratively refining prompts based on example animations. Finally, open-ended discussions probed perceptions of controllability, parameter tuning, interpretability, and creative agency.

\begin{figure}[h!]
    \centering
    \includegraphics[width=\linewidth]{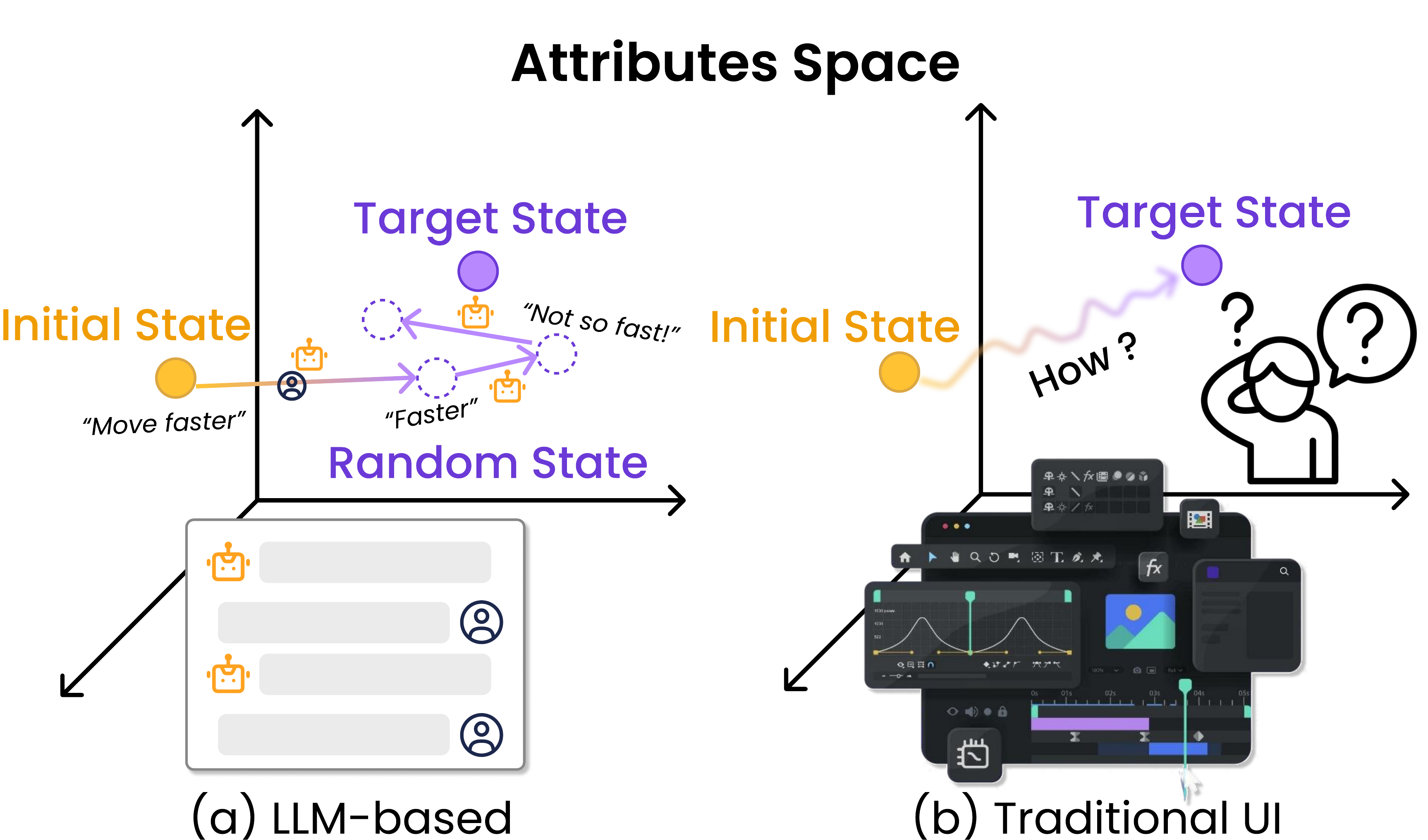}
    \caption{Limitations of current attribute control paradigms. (a) LLM-based prompting often produces unpredictable results and makes it difficult to precisely specify target attribute values. (b) Traditional UIs offer comprehensive control but rely on complex interfaces that are difficult for novice users to navigate. }
    \label{fig:attribute_space_limitation}
\end{figure}

\subsubsection{Challenges of Current Practices.}
\revise{As shown in Fig.~\ref{fig:attribute_space_limitation}, our formative study revealed complementary limitations in traditional authoring tools and LLM-based workflows:}

\begin{itemize}[leftmargin=5.5mm]
    \item \revise{Excessive low-level control: Conventional tools provide precise manipulation, but participants found their dense parameter spaces difficult to navigate. As one participant described, there were \q{too many widgets} to locate the relevant controls, increasing cognitive load during iterative refinement.}

    \item \revise{Obscured parameters: In LLM-based workflows, control became too abstract. Without access to underlying parameters, participants struggled to describe precise adjustments. As P3 noted, \q{I don’t know the current parameters, and it’s hard to describe the target change.}}

    \item \revise{Inefficient prompt iteration: Participants also found repeated prompting slow for minor changes. P5 explained, \q{I just want to slightly adjust the position, but I have to wait every time.} Manual code editing was possible, but locating the relevant variables in lengthy scripts remained cumbersome.}
\end{itemize}

\revise{Overall, traditional tools expose too many low-level controls, while LLM-based workflows hide them behind high-level prompts. This tension motivates interaction mechanisms for intuitive, in-situ adjustment of LLM-generated motion graphics while preserving generative flexibility.}

\subsection{Early Prototyping: Tech Probe}
Based on the insight from the previous study, we implemented a technical probe to explore the feasibility of in-situ refinement. The probe allowed users to first generate motion graphics and then dynamically overlay on-demand UI and interaction mechanisms to directly modify attributes. 

\subsubsection{Participants and Procedure}
To evaluate the effectiveness of this tech probe, we re-invited the six participants from Stage 1 (P1–P6), leveraging their established understanding of the study’s objectives. The session began with a guided walkthrough to familiarize participants with the interface’s core functionalities and the on-demand generation of control mechanisms. Following this orientation, participants engaged in a free exploration task where they were tasked with generating motion graphics and iteratively adjusting their attributes using the probe’s in-situ UI. 
We employed a think-aloud protocol to capture real-time feedback on usability hurdles and interaction nuances. 
The study concluded with a semi-structured interview where participants compared the tech probe with the manual and LLM-based practices identified in the previous stage, allowing us to pinpoint remaining limitations and further refine the design requirements for our final system.

\subsubsection{Tech Probe: The Baseline Workflow}
\begin{figure}
    \centering
    \includegraphics[width=\linewidth]{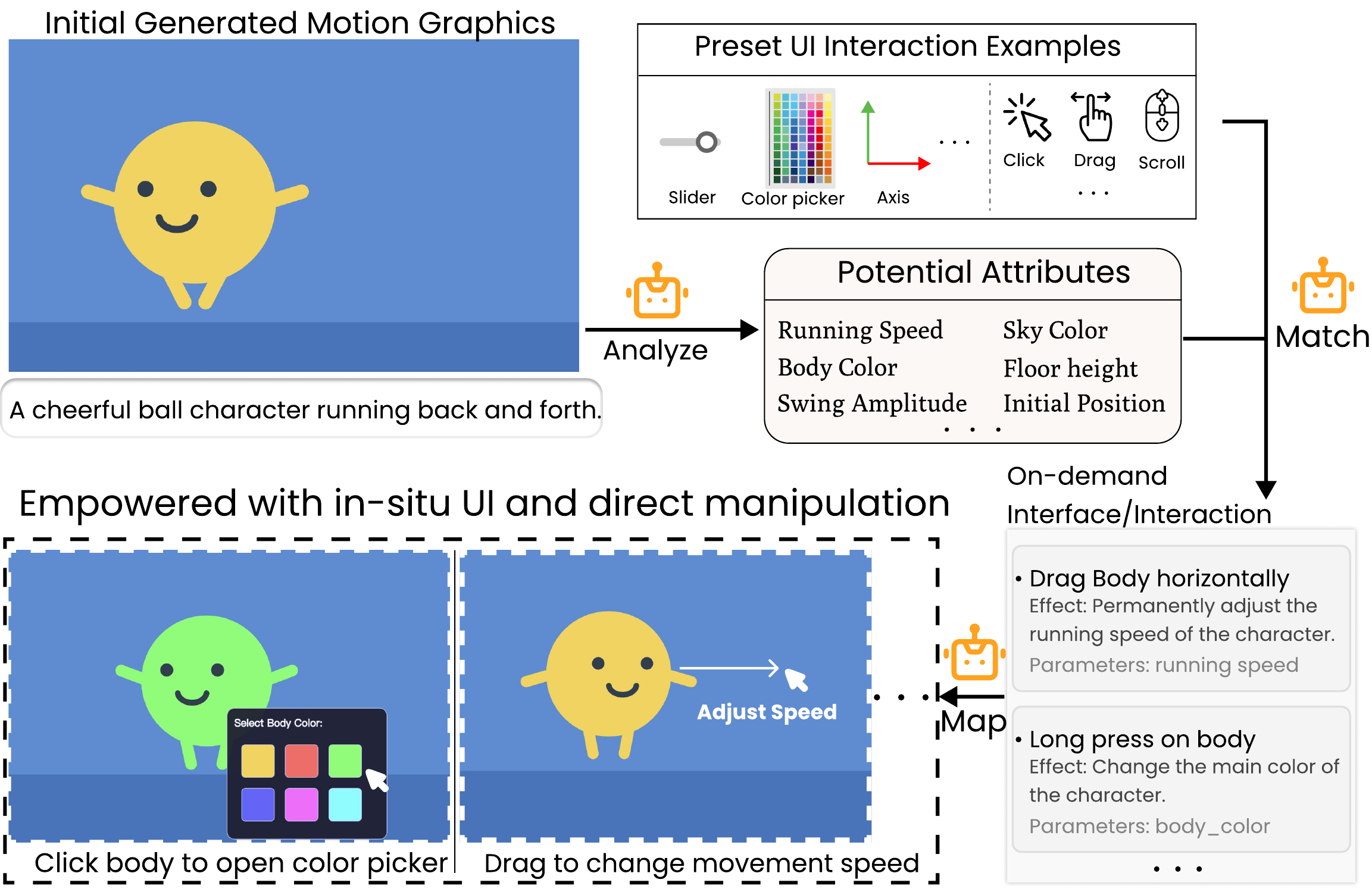}
    \caption{
    Pipeline of the tech probe for on-demand interaction generation. Given an initial motion graphic (top left), the system utilizes an LLM to (1) analyze the code context to identify potential attributes (e.g., speed, color); (2) match these attributes with preset UI interaction examples (e.g., sliders, drag gestures); and (3) map and inject these as in-situ interfaces directly onto the canvas. This allows users to perform real-time, fine-grained adjustments, such as dragging to change movement speed or clicking to open a color picker, directly on the visual elements.}
    \label{fig:tech_probe}
\end{figure}

The tech probe workflow begins with the user generating an initial motion graphic via a text prompt, followed by the core interaction phase: in-situ attribute adjustment (Fig.~\ref{fig:tech_probe}). 
After the initial result (i.e., the p5.js code and rendered animation) is generated, the LLM then examines the current motion graphics context, identifies potentially adjustable attributes, and determines appropriate direct manipulation strategies and corresponding UI components (e.g., object position can be mapped to drag-based interaction directly on the canvas).
We predefine a set of candidate UI elements and interaction patterns, which the LLM can selectively adapt, modify, or embed into the generated motion graphics code. Once this process is complete, users can directly manipulate visual elements on the canvas to adjust parameters in real time.
This workflow shifts refinement from repetitive, prompt-based regeneration to interaction within pre-constructed control scaffolds.
Rather than requiring users to iteratively re-prompt the LLM for every minor adjustment~\cite{masson2024directgpt,xie2025datawink}, we proactively expose a broad range of editable attributes through generated in-situ controls. 

\subsubsection{Positive Findings.}
Participants responded positively to the tech probe, especially the shift from static output to an interactive workspace. P4 described the transition from a \q{fixed animation} to an \q{interactive, editable} state as \q{very impressive}. P6 highlighted the natural feel of direct, on-canvas manipulation, while P2 appreciated the \q{on-demand} interface for staying \q{very clean} and avoiding \q{complex panels}. P5 found the generated interactions \q{very intuitive}, indicating effective mapping from attributes to control metaphors. Overall, these results suggest that pre-generated interaction affordances help bridge high-level generative intent and low-level control, enabling a more fluid workflow.

\subsubsection{Remaining Challenges}
Despite the encouraging feedback, the probe still presents limitations that hinder its transition toward more practical deployment. 

\textit{\textbf{Limited Discoverability of In-Situ Controls (C1).}}
Although attribute controls were designed to be intuitive, participants did not always immediately recognize how to manipulate specific parameters. For example, P6 remarked, \q{I only realized at the end that I could drag to change the circle’s radius.} Users often struggled to infer available controls without guidance, and P1 suggested, \q{It would be better if the system could directly tell me how to adjust it.} This uncertainty led participants to inspect interaction logic or consult documentation, disrupting the creative flow. These observations reveal a gap between generated and perceived affordances, indicating a need for contextual guidance or adaptive hints that reveal interaction possibilities at the moment of intent.


\textit{\textbf{Insufficiency in Control Resolution (C2).}}
Participants observed that the level of control detail did not always match their needs. Sometimes they wanted to make simple adjustments but were faced with overly complex interfaces, while at other times they needed fine-grained control that wasn’t available. For example, P6 said, \q{Sometimes I just want to tweak something quickly, but the interface becomes too complicated,} and P2 remarked, \q{Other times I need precise control, but the options are too limited.} Therefore, the desired control resolution for a given attribute varies across users and contexts, and it is important to account for this when generating on-demand interfaces.


\textit{\textbf{Limited Support for Collective Editing (C3).}}
Participants often wanted to manipulate visually or semantically related elements as a group. P2 noted, \q{I want to change them together,} and P3 added, \q{It would be nice to treat them as a group,} particularly when elements shared similar roles. However, current controls were object-centric, requiring adjustments like color or scale to be applied \q{one by one}. This design limits the efficiency of attribute control and highlights the need for mechanisms that can determine the appropriate editing scope without introducing excessive user interaction.


\textit{\textbf{Incomplete Coverage of Potential Controls (C4).}}
Although the system proactively generates commonly used attribute controls from the initial prompt and scene context, participants still encountered unsupported adjustments. The space of possible visual attributes and interaction mappings is inherently open-ended, making it impossible to pre-generate all potential controls.
For example, users wanted to \q{adjust the number of elements} (P1) or \q{change the width of the line} (P5), beyond the initially generated controls. This suggests that we need to provide an intuitive way that allows users to seamlessly supplement missing controls and dynamically expand the control boundaries.

\subsubsection{Design Guidelines}
Based on these challenges, we derive the following guidelines.

\textit{\textbf{Just-in-Time Hints for In-Situ Interaction (D1).}} 
Reveal  hidden controls through adaptive, in-situ hints triggered by user intent, guiding users on how attributes can be manipulated.

\textit{\textbf{Multi-Level Control Resolution (D2).}} 
Support on-demand expansion from coarse adjustments to fine-grained controls, balancing simplicity and precision.

\textit{\textbf{Semantic-Aware Collective Editing (D3).}} 
Enable simultaneous adjustment of semantically related elements, allowing users to operate on groups rather than individual objects.

\textit{\textbf{Complementary Control Expansion (D4).}} 
Allow dynamic extension of the control space, enabling users to introduce new parameters through lightweight interactions.
\begin{figure*}[ht!]
    \centering
    \includegraphics[width=\linewidth]{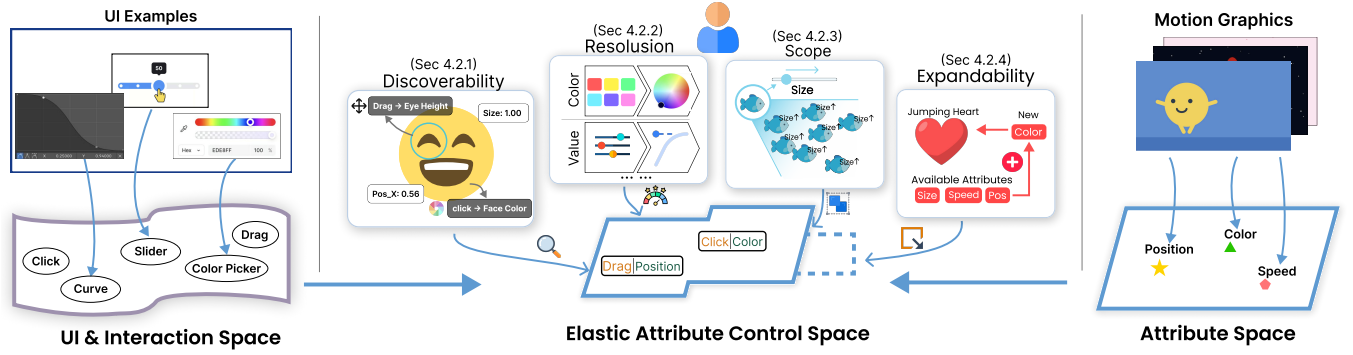}
    \caption{The framework of the Elastic Attribute Control Space. Spatula maps common UI examples and interaction modalities (left) to the underlying parameters of a motion graphic (right). The resulting space is "elastic," adapting across four key dimensions: (Sec 4.2.1) Discoverability for revealing affordances; (Sec 4.2.2) Resolution for multi-level precision; (Sec 4.2.3) Scope for collective editing; and (Sec 4.2.4) Expandability for on-demand parameter addition.}
    \label{fig:features}
\end{figure*}
\section{\tool: Elastic Attribute Control Space}
Drawing on these insights, we introduce \tool, an interactive authoring interface designed to support expressive and efficient attribute control.
\tool\ operates as a web-based, LLM-powered interactive canvas. Users generate motion graphics and iteratively refine parameters in real-time through on-demand generated direct manipulation and in-situ UI widgets.

\subsection{Overview}

Our task can be viewed as constructing an \textbf{\textit{Attribute Control Space}} for a given motion graphics instance. 
However, the attribute control space exposed in the technical probe is limited in several aspects: space discoverability (C1), control resolution (C2), control scope (C3), and space boundary (C4). 
Guided by the derived design implications, our goal is to transform this fixed control space into an \textit{elastic} one. To address these limitations, we transform this static scaffold into an \textit{elastic} space. 

We define an \textit{Elastic Attribute Control Space} as an adaptive and continuously reconfigurable interaction space, whose structure, granularity, and boundary can dynamically adjust in response to users’ intent and interaction context. Such a space should support easy exploration, accommodate varying levels of control precision, and expand or contract as needed.
To this end, we support exploration of the Attribute Control Space along four key directions:

\begin{itemize}[left=0pt]
    \item \textbf{Discoverability}: progressively revealing which attributes and interaction dimensions are available for manipulation.
    \item \textbf{Resolution}: providing different levels of control granularity for parameter refinement to match the required precision of a task.
    \item \textbf{Scope}: coordinating manipulation across multiple semantically related elements. 
    \item \textbf{Expandability}: dynamically extending the control space when existing parameters are insufficient.
\end{itemize}

The Elastic Attribute Control Space serves as the interactive bridge between high-level generative intent and low-level parameter adjustment.
Our research does not aim to build a full-fledged animation generation system. Instead, we focus on enabling on-demand parameter refinement over AI-generated results. 
Therefore, we propose a proof-of-concept system, while intentionally simplifying the full functionality of professional animation software (e.g., detailed timeline control, multi-layer sequencing, or advanced keyframing) following previous research~\cite{videocraft2025li,xia2023realitycanvas}.

\subsection{Exploring the Elastic Attribute Control Space}
As shown in Fig.~\ref{fig:features}, we introduce four strategies to support exploration of the attribute control space for more effective motion graphics editing.

\subsubsection{In-Situ Space Discovery.}
\begin{figure}[h!]
    \centering
    \includegraphics[width=\linewidth]{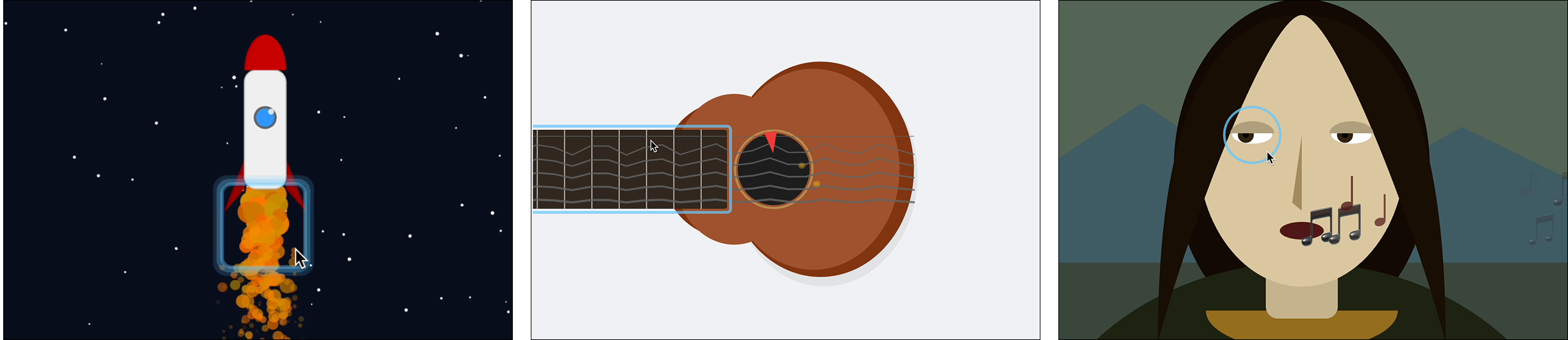}
    \caption{ Example Hints of what can be adjusted. Animatable components are highlighted with blue bounding boxes upon hovering. (Left) Exhaust in rocket flying animation. (Middle) The Neck in guitar played animation. (Right) The character's eye in singing animation.} 
    \label{fig:low-hint}
\end{figure}

A key challenge in on-demand attribute control is that the available attribute control space remains inherently latent, users cannot directly perceive the underlying interaction. To address this, we introduce \textit{in-situ space discovery} that progressively externalizes the attribute control space through contextual, canvas-embedded hints. These hints  are dynamically triggered based on user interaction signals (e.g., hover, focus, manipulation) and rendered as ephemeral overlays directly on the canvas.
To achieve this, we introduce progressive in-situ hints tailored to distinct moments of user intent (D1):

(1) \textit{What Can Be Adjusted.}  
At the exploration stage, users face uncertainty about which elements expose controllable attributes. To address this, the system offers an on-demand highlighting mechanism that selectively reveals interactable elements on the canvas (Fig. ~\ref{fig:low-hint}). 

\begin{figure}[h!]
    \centering
    \includegraphics[width=\linewidth]{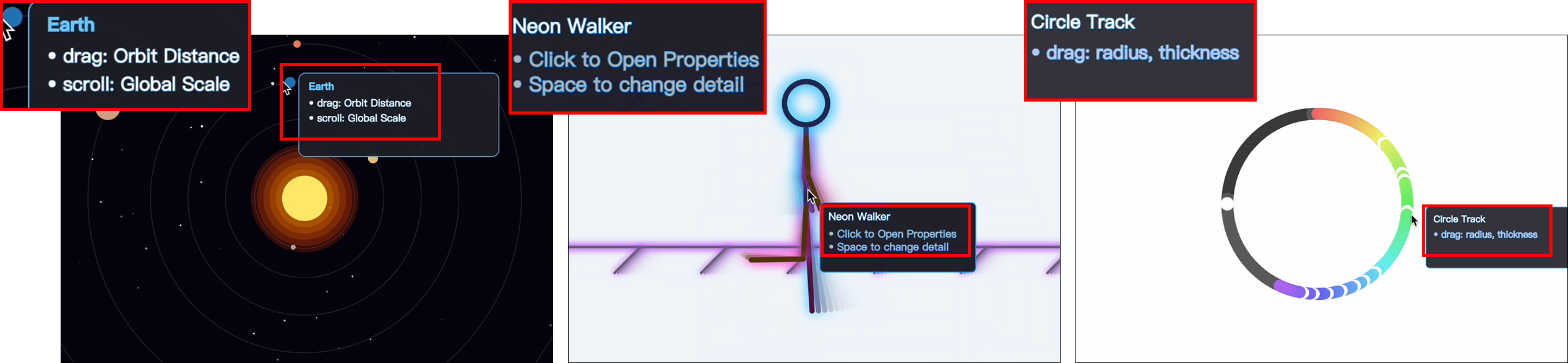}
    \caption{Example hints of how to adjust. On-hover tooltips specify available interactions and attributes for different animated elements: (Left) Earth in solar system animation. (Middle) Character in walking animation. (Right) The circle in loading spinner animation.} 
    \label{fig:middle-hint}
\end{figure}

(2) \textit{How to Adjust.}  
Once a target is identified, users need to understand the available manipulation strategies. Instead of requiring users to inspect external UI, the system generates localized interaction descriptors that are co-located with the element as users interact. These descriptors encode both the interaction modality (e.g., drag, click) and its semantic effect, enabling immediate action-to-attribute mapping within the visual context (Fig.~\ref{fig:middle-hint}).

\begin{figure}[h!]
    \centering
    \includegraphics[width=\linewidth]{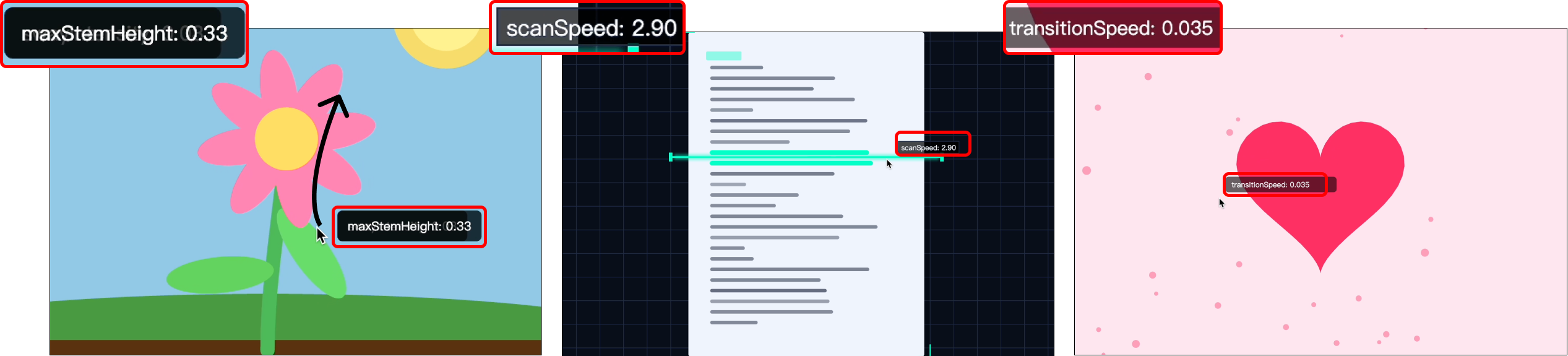}
    \caption{Detailed attribute values are revealed during interaction (e.g., dragging). (Left) Height value changes during a flower growing animation. (Middle) Scan speed adjustments in a file scan animation. (Right) Transition speed modifications in a heart morphing animation.} 
    \label{fig:high-hint}
\end{figure}

(3) \textit{What Has Been Adjusted.}  
During manipulation, we further expose the attribute space by revealing real-time parameter values alongside interaction.  The explicit value helps users interpret incremental changes to achieve fine-grained control that may be difficult to achieve through visual estimation alone, particularly in direct manipulation scenarios (e.g., horizontally dragging to adjust animation speed in Fig.~\ref{fig:high-hint}).

\subsubsection{Multi-Resolution Control}
We support attribute control resolution through \textit{elastic widget} (D2). Balancing interface simplicity with expressive precision has long been a challenge in interactive systems. Existing tools~\cite{ae,figmaColorWheel2026} typically address this by offering multiple, fixed control interfaces (e.g., sliders for coarse adjustment and curve editors for fine-grained tuning). However, users usually need to navigate across multiple menus and panels to access different levels of control. In contrast, we frame fidelity as a continuously expandable dimension within the attribute control space, rather than a set of predefined interface layers.

\begin{figure}[h!]
    \centering
    \includegraphics[width=\linewidth]{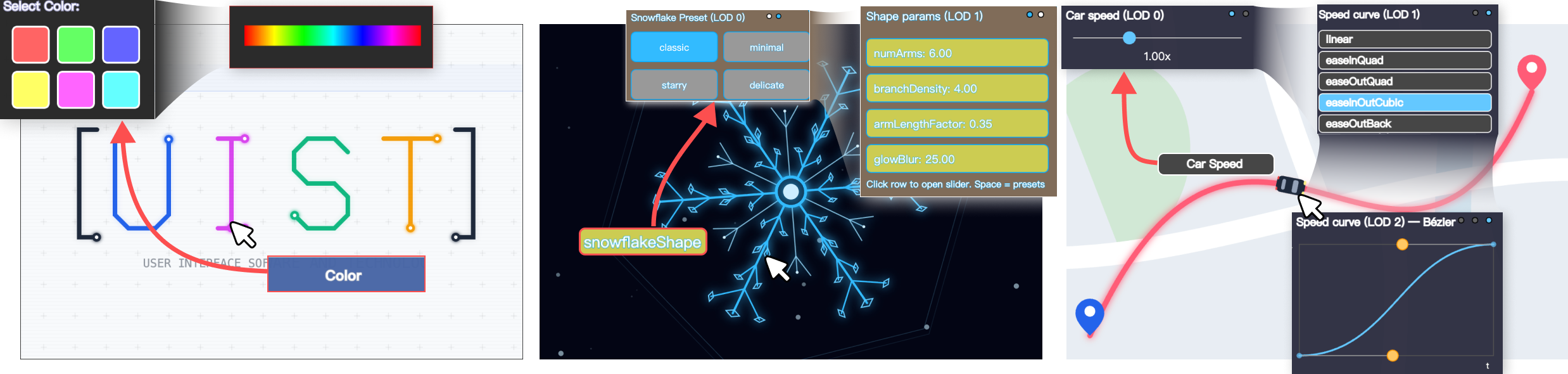}
    \caption{Examples of Multi-Resolution Control. (Left) color from palette to hue in the UIST Logo Animation. (Middle) shape from presets to procedural values in the Snowflake Animation. (Right) speed from slider to preset curves to Bézier curve for car navigation animation.}
    \label{fig:resolusion}
\end{figure}

Our system dynamically synthesizes a \textit{multi-resolution control widget} for each identified attribute. By default, users interact with low-resolution controls for rapid exploration. When higher precision is needed, the interface elastically expands in situ into a richer representation upon user invocation (e.g., from a scalar slider to a Bézier curve editor, or from discrete color palettes to a continuous HSV space in Fig.~\ref{fig:resolusion}).

\subsubsection{Semantic Scope Control}

\begin{figure}[h!]
    \centering
    \includegraphics[width=\linewidth]{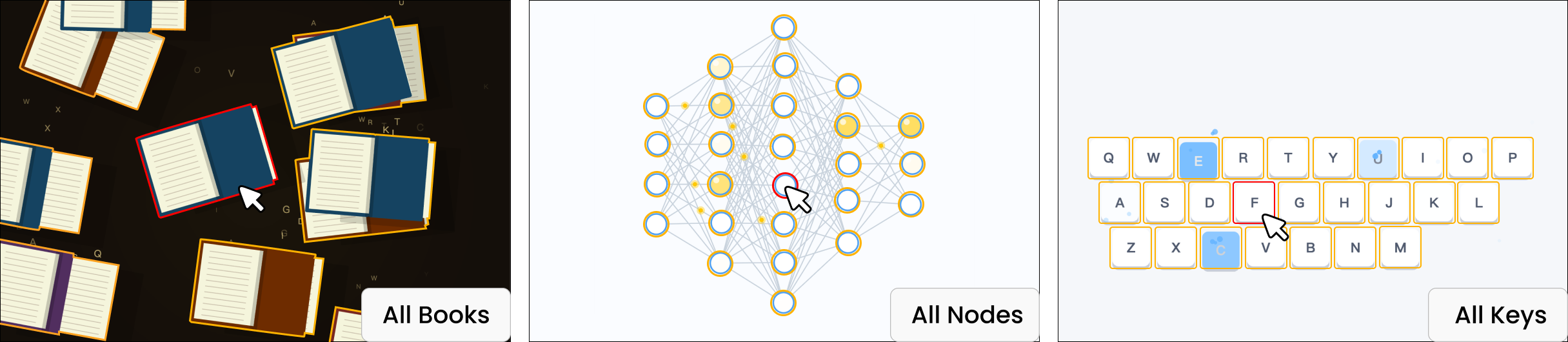}
    \caption{Examples of Scope Control. Selecting one element (red) extends the selection to semantically related elements (orange). In group mode, modifying the attributes of one element updates all elements in the group. (Left) Books floating. (Middle) Neural Network. (Right) Keyboard Animation.} 
    \label{fig:group}
\end{figure}

In many motion graphics scenes, multiple elements are semantically or visually related and are often perceived by users as a coherent group. For example, in neural network visualizations, many nodes share similar visual and functional properties (Fig. ~\ref{fig:group}-middle), and users may want to adjust them collectively rather than individually. 

To support this workflow, \tool~ provides a semantic group editing mechanism (D3), allowing users to manipulate attributes across a set of related elements simultaneously. The system identifies semantically similar elements based on their role, appearance, or functional category, and exposes collective controls for group-level adjustments. Users can still switch to individual editing for fine-grained refinement when needed, enabling seamless transitions between group-level and per-element manipulation.
\revise{Because these groups are inferred by the agent rather than fixed by the underlying structure, \tool~ supports multiple overlapping groupings for different editing goals, allowing the same element to participate in different groups.}

\subsubsection{Proactive Space Expansion}
\label{sec:expand}
\begin{figure}[h!]
    \centering
    \includegraphics[width=\linewidth]{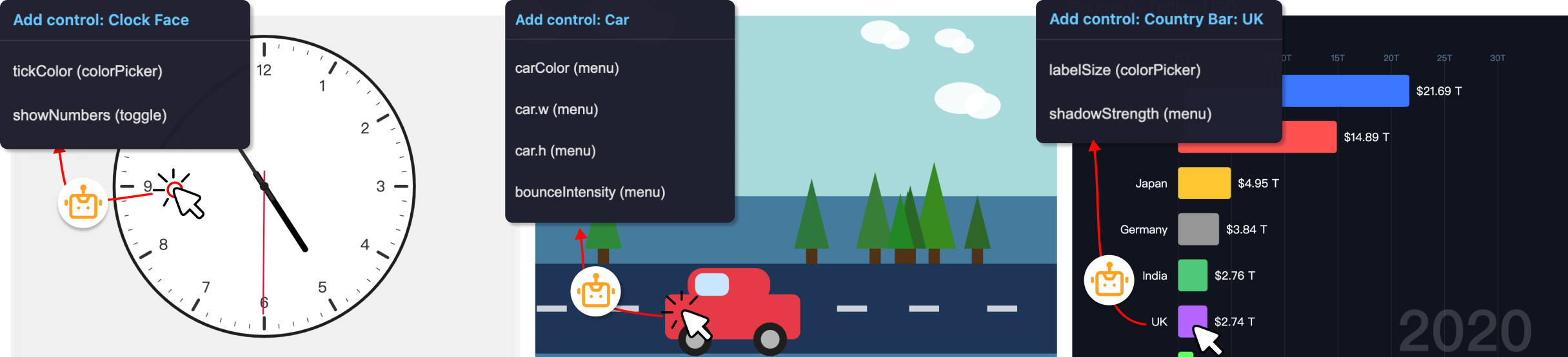}
    \caption{Examples of Attribute Space Expansion. (Left) clock face attributes in a clock animation. (Middle) car-related attributes in a driving animation. (Right) country bar (UK) attributes in a yearly GDP visualization animation.} 
    \label{fig:expand}
\end{figure}

To minimize user effort while maximizing creative possibilities, \tool~ implements proactive attribute expansion (D4). Inspired by prior proactive interface designs~\cite{zhao2026proactive,li2025designmemo}, the system initially suggests a c of potential attributes for each element (Fig.~\ref{fig:expand}). When users hover over an element and trigger the expansion, candidate attributes are presented, allowing users to selectively apply them to interactions.
In most cases, these suggestions are sufficient for constructing the desired interactions. When they are not, users can optionally provide lightweight textual input to describe missing controls to generate new UI elements on demand.
By allowing users to append new parameters to the existing scaffold, the system effectively manages the boundary of the attribute control space, ensuring it remains exhaustive without becoming overwhelming.

\subsection{Authoring Workflow}

\revise{We illustrate the authoring workflow through a usage scenario featuring Jack, a motion graphics designer. Jack has generated a space-themed animation in which an astronaut rides a horse through a field of planets and stars (Fig.~\ref{fig:teaser}). He is satisfied with the overall concept but wants to further refine its composition, appearance, and motion.}

\subsubsection*{\textbf{Step 1: Initialize Motion Graphics.}}
\revise{Jack begins by entering a prompt describing the animation. \tool\ generates the corresponding p5.js program and renders it on the canvas. He could also start by uploading an existing animation program if he wanted to continue editing a previous result.}

\subsubsection*{\textbf{Step 2: Identify Attributes and Map Interactions.}}
\revise{Jack selects [\texttt{Parse}] to make the animation editable. \tool\ analyzes the program, identifies candidate visual and motion attributes, and maps them to suitable interactions using the skill base. The resulting attribute--interaction mappings are displayed as structured cards, allowing Jack to inspect what the system has inferred without requiring him to configure each control manually.}

\subsubsection*{\textbf{Step 3: Apply Interactions.}}
\revise{Jack then selects [\texttt{Apply}]. \tool\ injects the inferred controls and event bindings into the running animation as a temporary interactive layer. The appearance of the animation remains unchanged, but Jack can now inspect and manipulate its elements directly on the canvas.}

\subsubsection*{\textbf{Step 4: Interact on the Canvas.}}
\revise{Jack first hovers over Saturn and sees an in-situ hint indicating that he can drag it to adjust its position (Fig.~\ref{fig:teaser}-a). He also finds controls for changing its radius and toggling its ring. As he manipulates these controls, the current values remain visible, helping him assess each adjustment.
Jack then decides that several planets should be edited consistently. He switches to group scope, and \tool\ highlights the related planets and applies his subsequent adjustment across the group (Fig.~\ref{fig:teaser}-b). To refine the astronaut's visor color, Jack begins with a color palette and expands it into a continuous hue control when he needs a more precise value (Fig.~\ref{fig:teaser}-c).
Finally, Jack wants to adjust the horse's body angle, but this control is not included in the initial scaffold. He invokes the expansion interface and selects \textit{Body Angle} from the suggested attributes. \tool\ adds a corresponding slider, allowing Jack to adjust the horse's posture directly on the canvas (Fig.~\ref{fig:teaser}-d).}

\subsubsection*{\textbf{Step 5: Save and Export.}}
\revise{After completing the refinement, Jack exports the updated animation. \tool\ removes the temporary authoring scaffold and writes the finalized values back to the original program. Alternatively, Jack can export the complete interactive package and retain the generated controls for later editing.}
\subsection{Implementation}
Our prototype is developed as a web-based authoring environment using React.js for the frontend and Node.js for the backend. The core of the interface features a centralized HTML5 Canvas rendered via the p5.js library, which serves as both the display for motion graphics and the primary surface for in-situ interaction. To enable intelligent UI generation, we utilize Gemini-3-Pro as the underlying LLM agent. Communication between the user’s canvas and the LLM is handled via asynchronous API calls, returning structured JSON metadata that defines the interactive bindings. The implementation of features are shown in Fig.~\ref{fig:implementation}, and more details are in the appendix. 

\subsubsection{Knowledge-Driven UI Synthesis.}
Our system embeds UI interactions for p5.js motion graphics using a three-layer schema: Attributes, Interaction Strategies, and UI Primitives. Attributes are editable parameters (e.g., position, size, opacity), interactions define how to manipulate them (e.g., drag, long-press), and UI primitives are concrete widgets (e.g., handles, sliders). 
\revise{Direct prompting often produces unstable attribute-control mappings and controls that do not align well with established authoring practices. To improve reliability, we built a skill base from a systematic survey of commercial authoring software, encoding over 300 UI controls, direct-manipulation techniques, attribute mappings, and interaction rules (see Appendix 9.1). The skill base determines the system’s ability to identify editable attributes and generate appropriate UI controls, and it can be further extended to support more complex editing cases.}
\revise{Given an executable animation, an Analyzing Agent uses the program structure and the skill base to identify common editable attributes. While some attributes may be already exposed as named variables, others are embedded in drawing expressions or animation logic. Spatula can expose these common cases by parameterizing the corresponding expressions and binding them to runtime controls, rather than requiring users to manually edit the code.}
An Applying Agent then selects appropriate UI widgets from the preset library based on the specification and injects them into the running animation as a temporary overlay, transforming it into an interactive workspace while preserving the original animation logic.

\subsubsection{Interaction Details.}
Through iterative refinement, we established a set of interaction rules to ensure seamless and intuitive user experiences (details are in appendix). These rules, encoded in the system’s knowledge base, guide the interpretation of p5.js animations and the synthesis of interaction logic through code injection. For instance, to avoid conflicts between click and long-press on the same element. The system consults these rules when generating interaction bindings.
Our system realizes interaction behaviors by injecting event-handling logic directly into the running p5.js code. Specifically, the system synthesizes native event listeners and lightweight detection routines based on the motion graphics context. These injected snippets monitor user inputs within the interactive canvas and connect to predefined JavaScript hooks that communicate with the external runtime system.
\revise{At runtime, the injected event handlers invoke predefined modules for highlighting, widget display, value updates, and group propagation, while UI rendering and state management remain in the external runtime. }When a shortcut is triggered, the injected detection logic invokes the corresponding hook (e.g., displaying a Hint UI overlay), while the actual UI rendering and state management are handled by pre-implemented runtime modules.

\begin{figure}[h!]
    \centering
    \includegraphics[width=1\linewidth]{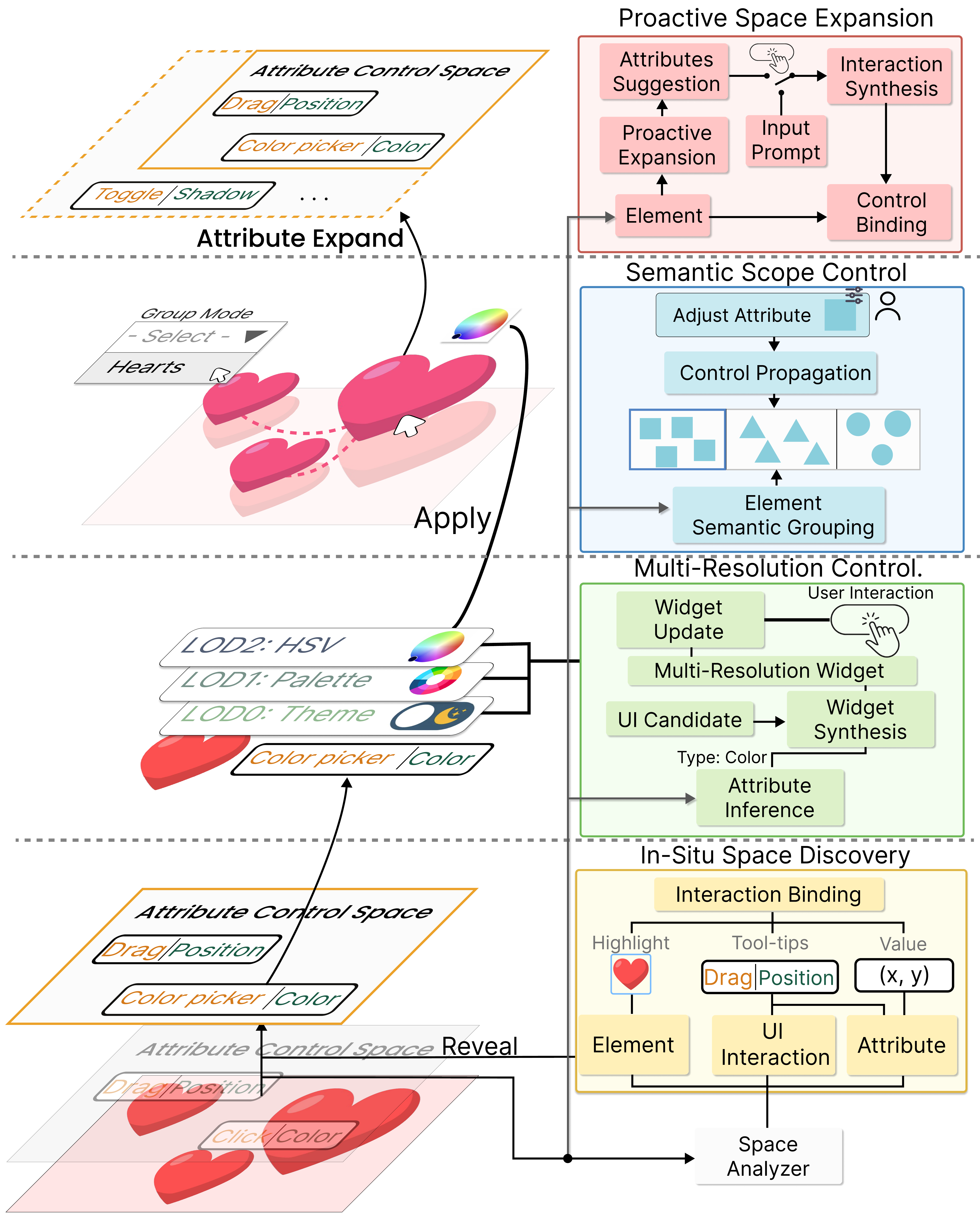}
    \caption{The pipeline of \tool. Powered by LLMs, the system constructs an attribute control space and supports user interaction through four key mechanisms for in-situ and progressive refinement of motion graphics.}
    \label{fig:implementation}
\end{figure}
\section{Evaluation}

\subsection{User Study}
We conducted a comparative user study to investigate how different interaction paradigms support exploration of the Attribute Control Space in motion graphics, with detailed refinement.

\subsubsection{Participants}

We recruited 12 participants (P1–P12) in local university communities. The group included five males and seven females with ages ranging from 23 to 33. To ensure a diverse range of expertise, we balanced the sample between six novices who had no formal training in motion graphics and six professional designers who possessed over two years of experience in motion graphics authoring. Among the professionals, four were regular users of Adobe After Effects, while two were in creative coding with p5.js and had already integrated LLM into their workflow. Compared with the formative study, which primarily involved experienced practitioners, this study intentionally includes novice users to better evaluate the accessibility and learnability of the proposed interface for first-time users. 

\subsubsection{Procedure}
The user study began with a 10-minute training session consisting of instructional videos and verbal explanations. To ensure a thorough understanding of the system, participants engaged in a free-exploration period to practice basic operations. A help button remained accessible on the interface to review interface functions at any time.

Following the training, we designed four comparative conditions across different interaction paradigms. These included 1) an integrated LLM interface for text-to-motion generation and iterative editing, 2) a separate panel adjustment interface to evaluate the necessity of in-situ control, 3) the tech probe interface from formative study, and finally 4) our proposed tool featuring the complete attribute control space. Participants interacted with the four interfaces sequentially in the order listed above in the following stages.

The evaluation was conducted in two stages. During the first stage, participants were presented with a specific motion graphic and a target goal. They were tasked with adjusting the animation to match the target as accurately as possible. 
In the second stage, an open-ended study allowed participants to either find or generate their own motion graphics and freely explore the control space to achieve self-defined creative goals. Each stage lasts approximately 40 minutes with four conditions, and each condition is repeated for two trials. \revise{Each trial refers to one editing task under one interface condition on one motion graphic, limited to 5min to observe how users quickly understand and interact with Spatula.}

After completing all tasks, participants provided subjective ratings on the expressiveness and usability of each interface and took part in a semi-structured interview to share qualitative feedback about their experience.

\begin{figure}[h!]
    \centering
    \includegraphics[width=1\linewidth]{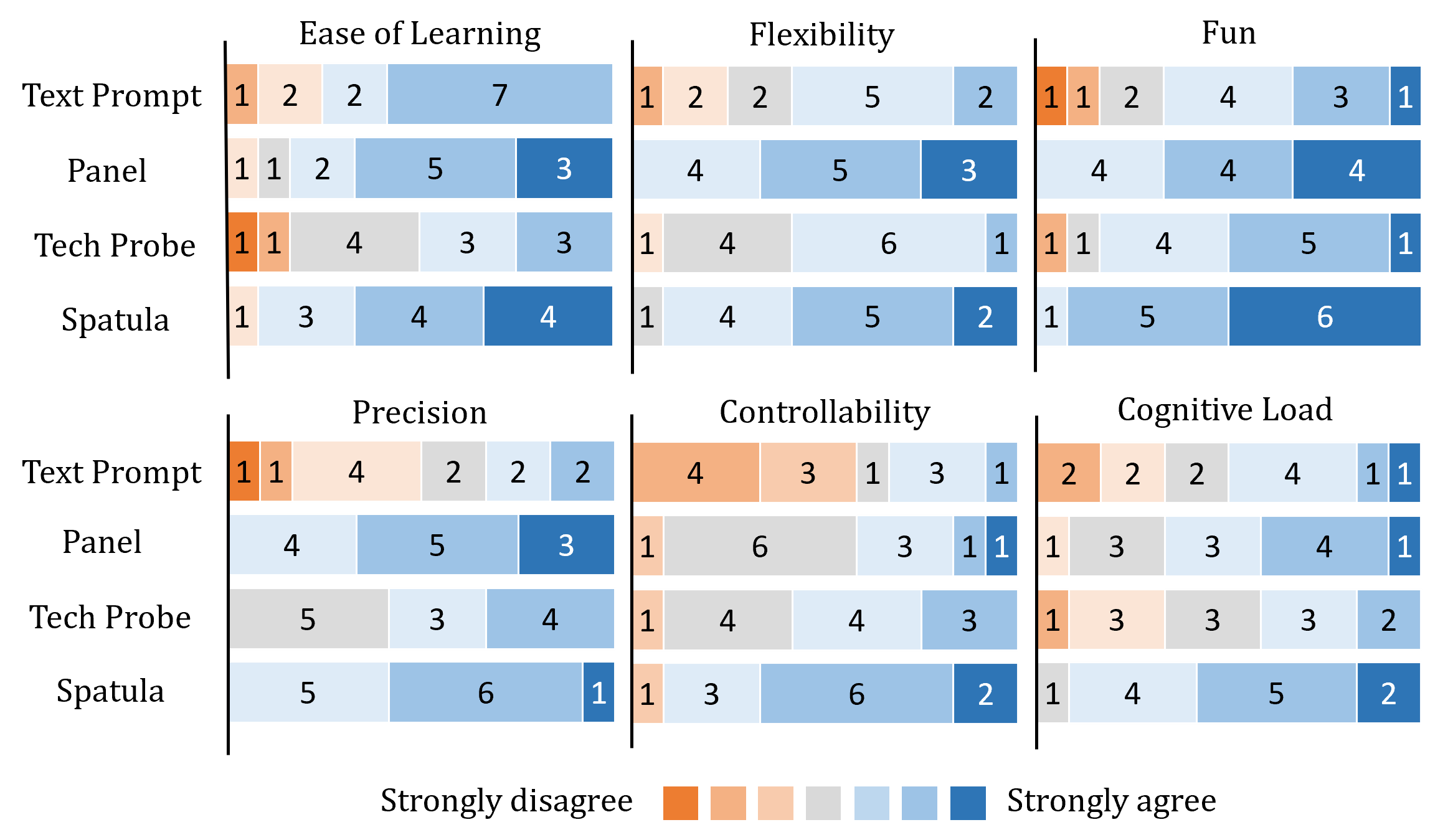}
    \caption{User ratings results from the user study.}
    \label{fig:results}
\end{figure}

\subsection{Results and Insights}
We report results and insights as shown in Fig.~\ref{fig:results}.

\textbf{\textit{Low-level manipulation enhances aesthetic engagement and creative exploration.}} The shift from high-level text prompts to granular attribute control changed how participants approached motion graphics. Compared with standard LLM workflows, where broad prompts leave aesthetic details to the model and often lead novices to passively accept ``good enough'' results, having a dedicated interface for exploring the attribute control space boosted engagement and creativity. Participants rated the system highly for controllability and enjoyment (Fig.~\ref{fig:results}), using the controls to fine-tune animations and embed their personal vision. This hands-on exploration often sparked new design ideas. As P3 noted, \q{Normally I just take whatever the AI gives me if it looks okay, but seeing all these exact parameters made me want to fine-tune the rhythm to match my own taste.} P7 added, \q{Playing around with the specific control curves actually gave me fresh inspiration for how the animation should feel, which I never would have thought of just by typing a prompt.}

\textbf{\textit{Exploring controls helped users understand the editable structure of animations.}}
\revise{Beyond completing individual edits, participants used the proposed dimensions to develop a working understanding of the control space. In the prompt-based baseline, the relationships among visual elements, editable attributes, and resulting changes remained largely implicit. With \tool, participants could discover controls in situ and observe how attribute changes affected the animation in real time. As P12 noted, the provided hints helped them \q{immediately understand how to edit.} Features like semantic scope control further made inferred relationships among elements visible, helping participants reason about whether an edit should apply to an individual element or a group. Together, these cues helped participants view the control space as an interconnected structure rather than isolated parameters.}

\textbf{\textit{Divergent interaction paths across expertise.}}
Elastic Attribute Control Space \revise{supported distinct interaction strategies across levels of expertise. In baseline conditions, novices struggled to map text prompts to visual outcomes}. With \tool, they exhibited a bottom-up, exploratory behavior. Relying on \textit{in-situ space discovery}, they used hover hints not just for navigation, but for brainstorming. \revise{Novices frequently invoked high-fidelity \textit{multi-resolution controls} (e.g., velocity curves) to explore how parameters affected the animation}, using immediate visual feedback to understand animation principles. P3 noted, \q{I couldn't describe the rhythm to the AI, but dragging the curve showed me exactly how the motion works.}
Conversely, experts adopted a goal-directed editing approach. During the open-ended exploration stage, rather than adjusting parameters indiscriminately, experts demonstrated a clear sense of intent and edited with high efficiency. They quickly grasped the available features and leveraged them purposefully, for instance, using \textit{semantic scope control} to rapidly block out scene-level changes. P10 noted, \q{These features are very intuitive to understand and easy to pick up.} When specific parameters were missing, experts \revise{used \textit{proactive space expansion} selectively}. P12 summarized, \q{I know exactly what I want to adjust, and I can just add the attribute if I need it.} Ultimately, while novices used the elastic space to expand their understanding, experts \revise{used it to reduce repetitive adjustments}.

\textbf{Complementary benefits of in-situ and separate UI interaction. }
Participants generally praised the intuitiveness of in-situ interaction (11 / 12), noting that it allows them to \q{edit wherever they want by directly clicking on the target} (P9). Such direct manipulation not only reduces the effort required to locate and adjust attributes, but also introduces a sense of playfulness into the creative process. For example, P3 described \q{dragging elements around just to explore how they look,} showing how in-situ interaction encourages open-ended exploration. Despite these advantages, we observed that separate panels offer complementary benefits in certain scenarios. In particular, P5 found them useful when in-situ controls occluded nearby elements or cluttered the visual workspace.  Therefore, on-demand interface generation could provide both in-situ and separate control interfaces to adapt to different contexts and user needs.

In conclusion, our evaluation does not aim to prove that \tool is universally superior to other paradigms. Instead, we demonstrate its unique advantages in parameter tuning and how it serves as a complementary approach to existing workflows.
\section{Discussion}

\subsection{Generative Interaction: From Fixed Toolbars to Interaction Compilers}

Beyond simply mapping sliders to variables, \tool\ represents a fundamental shift toward \textit{Generative Interaction}: a paradigm where the user interface is no longer a static, pre-designed artifact, but a transient, functional scaffold synthesized by an agent in response to user intent. Traditionally, adding a new control to a creative tool required a developer to manually define its UI, state management, and event listeners. By operationalizing the Elastic Attribute Control Space, we demonstrate that LLMs can act as ``interaction compilers'', translating high-level semantic intent into functional, in-situ code snippets on the fly. This suggests a future where authoring tools are no longer defined by their fixed toolbars, but by their ability to dynamically expand their interaction surface area to match the unique procedural depth of any generated object.

Current AI-assisted authoring often forces a trade-off between the ``magic'' of zero-shot generation and the ``control"'' of manual manipulation. \tool~ introduces a middle ground where the LLM’s role shifts from a mere content generator to a UI architect. We demonstrate that the black-box nature of LLM outputs can be externalized into a series of interactive scaffolds that bridge the gap between prompting and direct manipulation. This paradigm suggests a future for malleable AI, where every generated objects, such as a motion graphic, a 3D model, or a snippet of code, can have a built-in, on-demand interface tailored to specific latent attribute. 

\subsection{Applications}
\label{sec:application}

While we focus on motion graphics to explore how on-demand in-situ interfaces support attribute adjustment in LLM-based workflows, the design generalizes to a wide range of authoring tasks. We view our approach as a modular, plug-and-play component where the specific design context can be swapped to suit different domains. To demonstrate this versatility, we implemented two additional proof-of-concept cases.

\begin{figure}[h!]
    \centering
    \includegraphics[width=\linewidth]{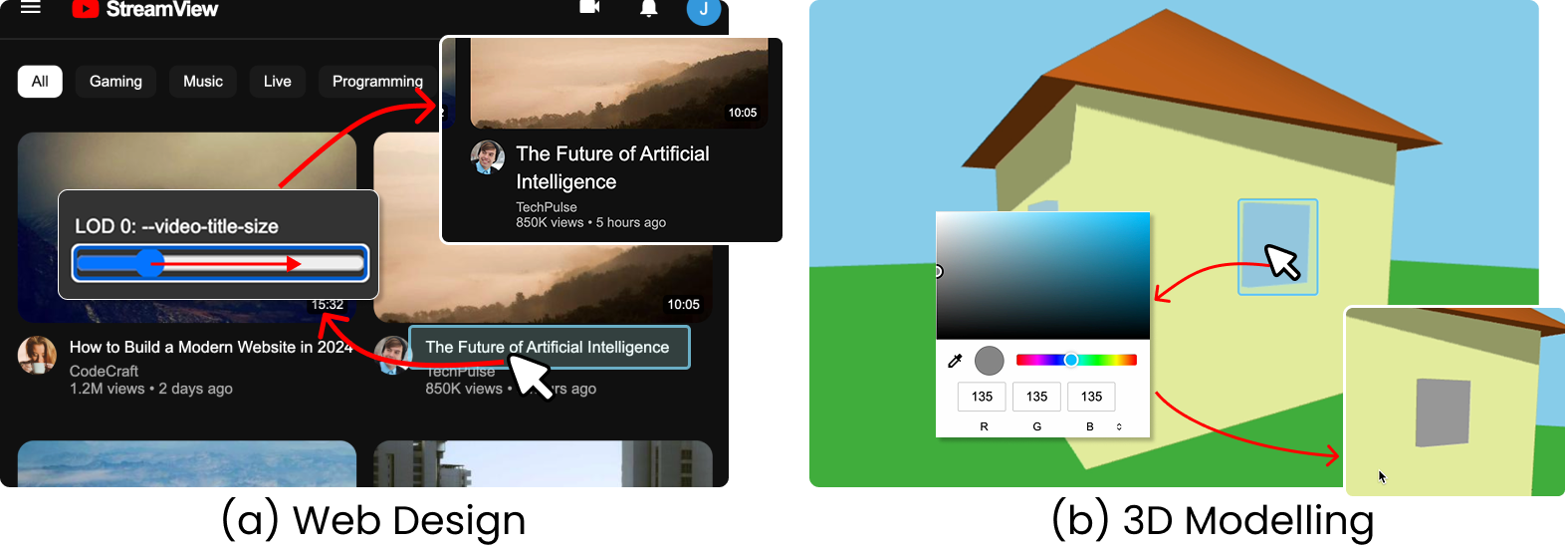}
    \caption{Application examples of \tool. (Left) In web design, users adjust the CSS style of a title (e.g., size). (Right) In 3D house modeling, users modify window color.} 
    \label{fig:application}
\end{figure}

\textbf{Web Design}. In this application (Fig~\ref{fig:application}-left), we allows users to manipulate CSS properties directly on rendered elements. The LLM generates layout or styling controls exactly where the cursor interacts with the webpage, enabling direct visual adjustments.

\textbf{3D Modeling}. Similarly, the application (Fig~\ref{fig:application}-right) analyzes geometry scripts to expose transformation handles or material sliders directly on 3D meshes. By identifying the underlying parameters of the 3D objects, the interface provides localized tools for spatial manipulation without requiring complex menu navigation.

Beyond these examples, we envision this interaction paradigm extending to other complex fields such as video editing~\cite{liu2024generativevideopropagation} or architectural design integrate with the embedding of the attribute control signal into a generative model.

\section{Limitations and Future Work}
\textit{\textbf{Overwhelming Attribute Spaces in Complex Contexts. }}
Although we aim to extract enough primary attributes (with technical evaluation in Appendix~\ref{sec:tech_eval}), in a more complex context (e.g., intricate motion graphics or web design in Sec.~\ref{sec:application}), the underlying attribute space can become extremely large. Generating interaction affordances for all these attributes can be time-consuming when applying interactions to the context, and more importantly, can degrade the LLM’s performance as in long context~\cite{liu-etal-2024-longgenbench}. In addition, the system may struggle to identify a complete set of relevant attributes, or fail to correctly implement the corresponding control logic. While our current implementation mitigates this by generating only major attributes and expanding controls on demand (Sec.~\ref{sec:expand}), it mainly reduces UI density rather than solving attribute discovery. Identifying and selecting appropriate attributes for control interactions remains a challenge and requires further investigation.

\textit{\textbf{Adaptive Interface for Attribute Control.}}                         
A promising yet challenging direction is shifting from explicit invocation to proactive intent prediction. In principle, the system could anticipate which attributes users will adjust~\cite{shaikh2025creating}, highlight relevant elements, and adapt control scope or resolution based on behavioral signals (e.g., action history, cursor trajectory). However, our prototypes show that user intent is highly variable and non-deterministic, making reliable prediction difficult. We therefore prioritize reliability with an explicit invocation model, where users trigger in-situ interfaces via cursor actions or hotkeys. Still, leveraging behavioral signals for more robust intent prediction remains an important direction for future work.

\textit{\textbf{The Boundary between Attribute Tuning and Content Restructuring.}} 
The distinction between an ``attribute'' and a fundamental content restructuring is often ambiguous.
\revise{For example, in Fig.~\ref{fig:resolusion} (middle), the snowflake shape is generated from procedural parameters and can therefore be exposed as an adjustable attribute. In contrast, the flower shape in Fig.~\ref{fig:high-hint} (left) is hard-coded as drawing commands, so making it editable would first require a different underlying representation. Similar boundaries arise in more complex edits: geometry-level control may require parameterized shape representations (e.g., SDFs), while choreography-level tuning may require explicit relations among elements, such as staggered timing or coordinated motion. Spatula's expansion mechanism can partially support these cases when users request additional motion-related controls, but fully inferring such structures remains a trade-off between automation reliability and manual specification effort.}
\revise{In \tool, this practical boundary is largely determined by the skill base: an aspect of the animation can be treated as an editable attribute when corresponding mappings, interaction rules, and parameterization strategies are available. As the skill base grows to cover richer representations and relations, this boundary can expand to support more complex forms of editing.}

\section{Conclusion}

We introduce \tool, a web-based proof-of-concept system for generating on-demand, in-situ interfaces and interactions for attribute control. Focusing on motion graphics, we begin with a tech probe and propose four connected design strategies to support diverse user needs: in-situ space discovery, multi-resolution control, semantic scope control, and proactive space expansion.
Through a comparative user study, we not only show that \tool effectively supports users, but also observe diverse interaction patterns across different users. Our application demonstrations further suggest that the design is feasible across creative contexts, such as web design and 3D modeling. We envision a shift toward generative interaction, where users actively steer and refine generated content, maintaining control through the process.


\bibliographystyle{ACM-Reference-Format}
\bibliography{reference}

\section{Appendix}

\subsection{UI Knowledge Base Examples}
\begin{figure*}[ht!]
    \centering
    \includegraphics[width=\linewidth]{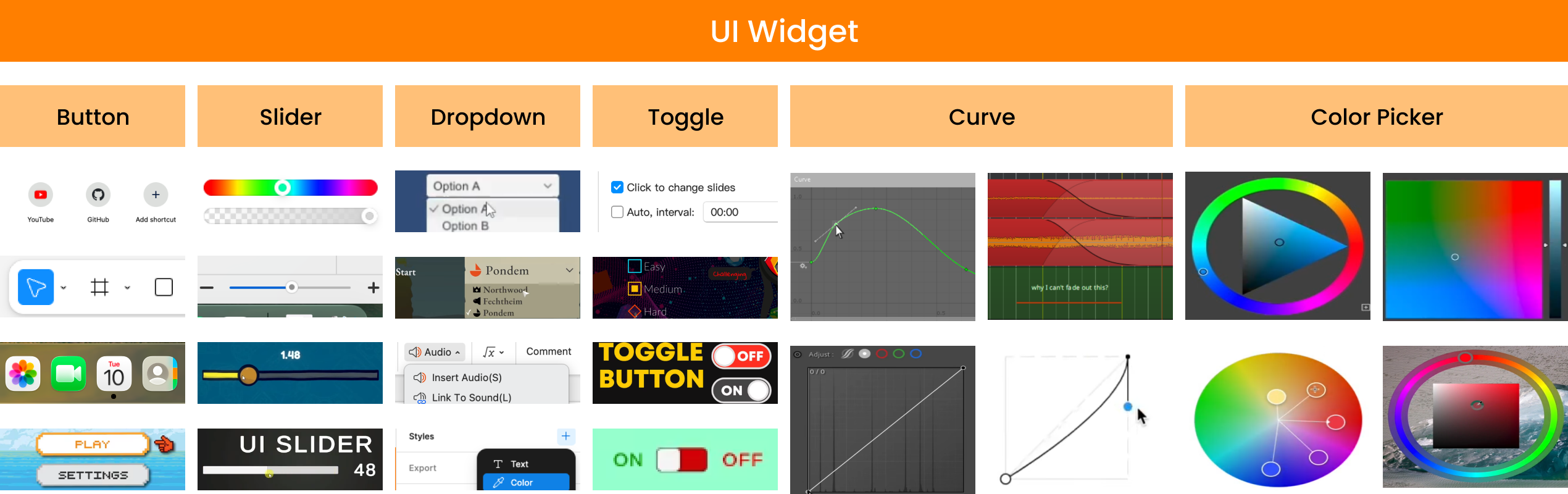}
    \caption{UI Widget Examples from commercial tools.}
    \label{fig:UI_widget}
\end{figure*}

We constructed a knowledge base linking motion graphics attributes to their corresponding user interface elements and interactions. To build this, we systematically surveyed currently available commercial software (Fig.~\ref{fig:UI_widget}) and widely used interaction primitives (Table~\ref{tab:interaction_primitives}). We prioritized commonly adopted controls that are directly relevant to our prototype, selecting examples that illustrate typical attribute manipulation patterns. While many additional interfaces and interactions exist (e.g., timeline scrubbing, nested parameter panels, or procedural node graphs), we selected only commonly used controls for our prototype. This choice reduces system complexity and focuses on core attribute manipulation tasks.

\begin{table}[htbp]
\centering
\caption{Interaction Primitive Design Space}
\label{tab:interaction_primitives}
\renewcommand{\arraystretch}{1.3} 
\begin{tabular*}{\columnwidth}{@{\extracolsep{\fill}} l l l @{}}
\toprule
\textbf{Dimension} & \textbf{Mouse} & \textbf{Keyboard} \\
\midrule
\textbf{Discrete Trigger} 
& \begin{tabular}{@{}l@{}}\textbullet~Click \\ \textbullet~Double Click \\ \textbullet~Right Click\end{tabular} 
& \begin{tabular}{@{}l@{}}\textbullet~Key Press \\ \textbullet~Key Sequence\end{tabular} \\
\midrule
\textbf{Continuous 1D}    
& \begin{tabular}{@{}l@{}}\textbullet~Scroll \\ \textbullet~Press and hold\end{tabular} 
& \begin{tabular}{@{}l@{}}\textbullet~Press and hold \\ \textbullet~Key Shortcut\end{tabular} \\
\midrule
\textbf{Continuous 2D}    
& \begin{tabular}{@{}l@{}}\textbullet~Point to \\ \textbullet~Drag\end{tabular} 
& \multicolumn{1}{c}{None} \\ 
\bottomrule
\end{tabular*}
\end{table}

\begin{table*}[t]
\centering
\caption{User Interface Example Sources. In Figures 14, we have used user interface example from videos published by the following creators on YouTube and some companies.}
\label{tab:youtube_sources}
\renewcommand{\arraystretch}{1.5} 

\begin{tabular*}{\textwidth}{@{\extracolsep{\fill}} l p{12cm} @{}}
\hline 
\textbf{Copyright Name} & \textbf{Channel Link} \\
\hline
© Copyright Google & \url{https://www.google.com/} \\
© Copyright Figma  & \url{https://www.figma.com/} \\
© Copyright Mac  & \url{https://www.apple.com/mac/} \\
© Copyright AIA & \url{https://www.youtube.com/watch?v=IuuKUaZQiSU&t=1395s} \\
© Copyright PowerPoint & \url{https://www.microsoft.com/en-us/microsoft-365/powerpoint} \\
© Copyright Tarodev & \url{https://www.youtube.com/watch?v=nTLgzvklgU8} \\
© Copyright SpeedTutor & \url{https://www.youtube.com/watch?v=oya8_SlLXb0} \\
© Copyright Driple Studios & \url{https://www.youtube.com/watch?v=a_vDunGhvRw&t=292s} \\
© Copyright Christina Creates Games & \url{https://www.youtube.com/watch?v=E9AWlbPGi_4} \\
© Copyright Grafik Games & \url{https://www.youtube.com/watch?v=vhfzpKWWA-A} \\
© Copyright Unity & \url{https://docs.unity3d.com/Manual/EditingCurves.html} \\
© Copyright DAW & \url{https://discourse.ardour.org/t/automation-curves-part-ii/107405} \\
© Copyright Adobe & \url{https://helpx.adobe.com/photoshop/using/curves-adjustment.html} \\
© Copyright Adobe & \url{https://color.adobe.com/create/color-wheel} \\
© Copyright Blender & \url{https://docs.blender.org/manual/en/latest/interface/controls/templates/color_picker.html} \\

\hline
\end{tabular*}
\end{table*}

\subsection{Implementation of Interaction Features}

To realize the elastic attribute control space, we implement the four core interaction features as dynamic runtime injections and state-driven widget management over the canvas. Below we detail the algorithmic formulation and implementation logic for each feature.

\subsubsection{In-Situ Space Discovery}
The discovery mechanism dynamically externalizes latent attributes through context-aware spatial detection. Let $\mathcal{E}$ be the set of interactive elements in the motion graphic. During the code synthesis phase, the system injects lightweight bounding geometry tracking for each element $e \in \mathcal{E}$. 

At runtime, given the pointer coordinates $p = (x, y)$, the system constantly evaluates a hit-testing function $H(e, p)$. If $H(e, p) = \text{true}$, the element enters a \textit{hovered} state. The system then queries the interaction schema to retrieve the mapped attribute $A_e$ and interaction modality $M_e$ (e.g., drag, click), rendering an ephemeral bounding box and a tooltip. During active manipulation, an event listener hooks into the parameter update stream, continuously extracting the real-time value $v_t$ of $A_e$ and rendering it as a floating label anchored near $p$, providing immediate quantitative feedback.

\subsubsection{Multi-Resolution Control}
We formulate multi-resolution control as a state machine of widget representations. Each editable attribute $A$ is associated with an ordered set of Level-of-Detail (LOD) widgets $\mathcal{W}_A = [W_0, W_1, \dots, W_k]$ ranked by granularity (e.g., discrete color palette $\rightarrow$ continuous hue slider).

Let $l \in [0, k]$ be the current LOD index. When the user triggers the LOD switch (e.g., via the Space key), the system executes a state transition:
\begin{equation}
    l_{new} = (l + 1) \bmod |\mathcal{W}_A|
\end{equation}
The system immediately unmounts the current widget $W_{l}$ and mounts $W_{l_{new}}$. Crucially, to maintain visual continuity, the current parameter value $v$ is projected into the new widget's state space via a mapping function $v \rightarrow W_{l_{new}}(v)$. This allows seamless switching between coarse exploration and fine-grained tuning without losing the current editing context.

\subsubsection{Semantic Scope Control}
To support collective editing, the system identifies semantically related elements during the static analysis phase. Elements instantiated from the same class, array, or functional role are clustered into a semantic group $\mathcal{G} = \{e_1, e_2, \dots, e_n\}$. 

When Group Edit Mode is activated, the system overrides the default object-centric update logic. Let $v'$ be the new attribute value applied to an anchor element $e_i \in \mathcal{G}$. The system intercepts the update event and broadcasts it across the entire group using a propagation function:
\begin{equation}
    \forall e_j \in \mathcal{G}, \quad e_j.A \leftarrow v' \quad \text{(or } e_j.A \leftarrow e_j.A + \Delta v \text{)}
\end{equation}
At the code level, this is achieved by dynamically traversing the underlying data structure array in the runtime environment (e.g., \texttt{window[group.elementType]}) and applying the property modification iteratively. This ensures synchronized visual updates across all grouped elements in real-time.

\subsubsection{Proactive Space Expansion}
The proactive expansion feature dynamically augments the control space on demand. Let $\mathcal{P}_e$ be the exhaustive set of potential attributes for an element $e$ inferred by the LLM, and $C_e \subset \mathcal{P}_e$ be the subset of currently implemented controls.

When the user invokes the expansion trigger while hovering over $e$, the system computes the available candidate set:
\begin{equation}
    \mathcal{P}_{avail} = \mathcal{P}_e \setminus C_e
\end{equation}
An in-situ contextual menu is then rendered to display $\mathcal{P}_{avail}$. Upon the user's selection of a new attribute $A_{new} \in \mathcal{P}_{avail}$, the system dynamically synthesizes the corresponding interaction binding code (AST injection) for $A_{new}$. The modified code is seamlessly re-evaluated into the runtime, instantly exposing the new widget and permanently expanding the attribute control space.

\subsection{Interaction Rules}
We present a set of example interaction rules that guide the agent. These are simplified illustrations derived from iterative development and formative study feedback. In practice, the full rules are more comprehensive, following a structured template and accompanied by input–output examples.
\label{app:interaction-rules}

\subsubsection{General Principles}
\begin{itemize}
  \item \textbf{Preserve existing behavior.} Interaction injection must not alter the original visual/functional behavior unless explicitly requested.
  \item \textbf{Temporary widgets, permanent effects.} UI widgets (menus, sliders, pickers, overlays) must appear only during interaction and disappear after an action; the resulting change must persist as a parameter/state modification rather than a transient visual effect.
  \item \textbf{Non-intrusive augmentation.} Added interaction layers should be lightweight and should not introduce unnecessary modes or complexity beyond what is needed for the requested controls.
\end{itemize}

\subsubsection{Gesture Conflict Resolution (Click / Long-press / Drag)}
\begin{itemize}
  \item \textbf{No early commitment.} When multiple gestures apply to the same target, the system must not immediately commit to a click or open a menu at press-down time.
  \item \textbf{Drag cancels click/long-press.} If movement exceeds a small spatial threshold, treat the action as drag and cancel any pending click/long-press recognition.
  \item \textbf{Long-press requires stillness.} Long-press is triggered only if press duration exceeds a time threshold and the pointer remains within a small movement bound.
  \item \textbf{Click is short-press without drag.} A click is recognized only when the press duration is below the long-press threshold and no drag was detected.
\end{itemize}

\subsubsection{Widget Visibility and Dismissal}
\begin{itemize}
  \item \textbf{Outside-click dismissal.} Any open widget must close when the user clicks/taps outside all widgets.
  \item \textbf{Escape-to-close.} A global cancel action must be provided to close all open widgets/menus immediately.
  \item \textbf{Auto-close after apply.} After a parameter change is applied via a widget, the widget should close promptly to reinforce its temporary nature.
\end{itemize}

\subsubsection{Progressive Guidance (Three-Stage Interaction Support)}
\begin{itemize}
  \item \textbf{Stage 1: Discoverability (hover).} Provide lightweight cues that indicate an element is editable without disrupting the ongoing animation.
  \item \textbf{Stage 2: How-to guidance (engagement).} Upon explicit engagement, provide clearer guidance about the available manipulation method(s).
  \item \textbf{Stage 3: Value feedback (mandatory).} While adjusting parameters, show contextual value feedback that remains readable, avoids occlusion, and auto-hides after inactivity.
\end{itemize}

\subsubsection{Level-of-Detail (LOD) Controls}
\begin{itemize}
  \item \textbf{Multi-LOD editing for complex parameters.} Complex parameters should support multiple control granularities (e.g., simple vs. advanced).
  \item \textbf{Explicit LOD switching.} Provide a consistent, discoverable method to switch LOD levels during editing, with a visible hint indicating how to switch.
\end{itemize}

\subsubsection{Expandable Control Space }
\begin{itemize}
  \item \textbf{On-demand expansion.} When the user requests expansion (e.g., via a dedicated expansion trigger), present a short menu of not-yet-exposed candidate controls for the currently focused element.
  \item \textbf{Sufficient candidate breadth.} The system should ensure each element type has multiple plausible adjustable parameters so expansion remains useful over time.
\end{itemize}

\subsubsection{Global Shortcuts and Navigation}
\begin{itemize}
  \item \textbf{Highlight-on-demand.} Visual highlighting of editable targets should be strictly gated by an explicit user request (to avoid persistent visual clutter).
  \item \textbf{Operation instructions on demand.} A dedicated shortcut should display concise operation instructions near the pointer only while held/active.
  \item \textbf{Optional pause-for-precision.} Provide an optional interaction mode that pauses motion to facilitate precise parameter editing, with a clear indicator when active.
  \item \textbf{Cycling through targets.} Provide a way to cycle across interactive elements when fine-grained selection is needed without introducing a heavy selection mode.
\end{itemize}

\section{Technical Evaluation}
\label{sec:tech_eval}
The efficacy of our on-demand control generation fundamentally depends on the LLM's capability to interpret motion graphics code (p5.js) and accurately extract meaningful semantic attributes. In this section, we quantitatively evaluate this attribute extraction task, primarily testing our default model, Gemini 3, and benchmarking its performance against other contemporary LLMs (e.g., GPT-5, Kimi-2.5, Qwen).

\subsubsection{Dataset Preparation} To facilitate a rigorous evaluation, we constructed an attribute extraction dataset consisting of 50 unique p5.js motion graphics scripts. For UI control purposes, we structured the extracted parameters into a two-tier hierarchy: \textit{Primary Attributes:} Essential parameters that fundamentally govern the animation's visual aesthetics and motion dynamics (e.g., global speed, quantity, main color schemes). These represent the most frequent adjustments desired by users. \textit{Secondary Attributes:} Auxiliary parameters used for subtle refinements (e.g., stroke weights, secondary opacities, minor physics offsets).

\begin{figure}[h]
    \centering
    \includegraphics[width=1\linewidth]{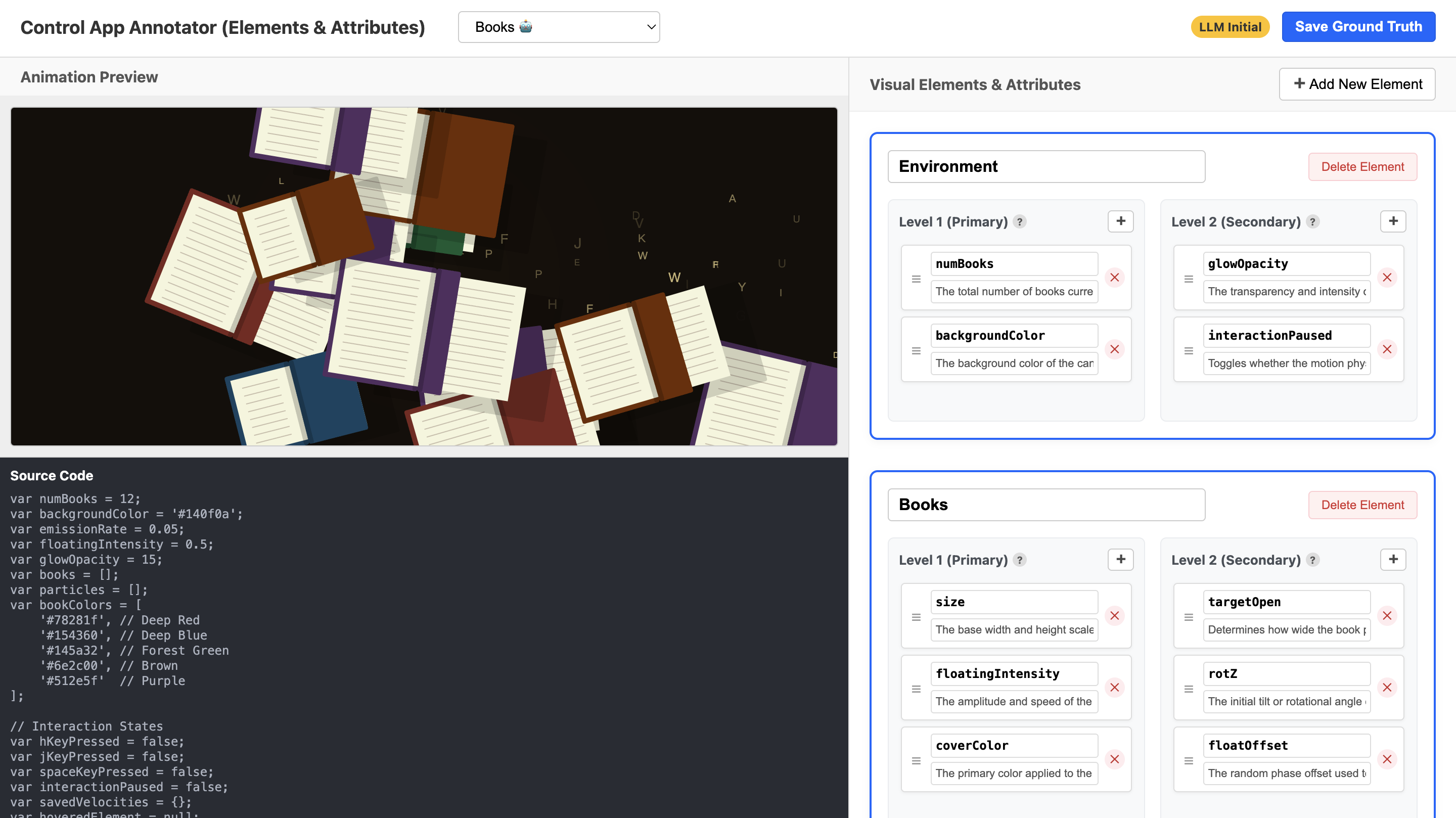}
    \caption{Interface for annotating motion graphics attributes to construct the ground-truth dataset.}
    \label{fig:datasets_interface}
\end{figure}

The base motion graphics scripts were sourced from the design artifacts produced during our formative study, supplemented by additional LLM-generated examples to ensure diversity. Given the subjective nature of determining meaningful UI controls, we established our ground truth using a human-in-the-loop annotation pipeline. First, we utilized an LLM to automatically generate an initial set of primary and secondary attributes for each script. Subsequently, we recruited 5 expert motion graphics designers to act as annotators. Using a custom annotation interface (Fig.~\ref{fig:datasets_interface}), the experts systematically reviewed the LLM's initial predictions, re-categorizing misaligned attributes, and manually appending any overlooked parameters. The finalized, expert-validated attribute sets were then adopted as the ground truth for our benchmark. In addition, it also could be used

\begin{figure}[h]
    \centering
    \includegraphics[width=1\linewidth]{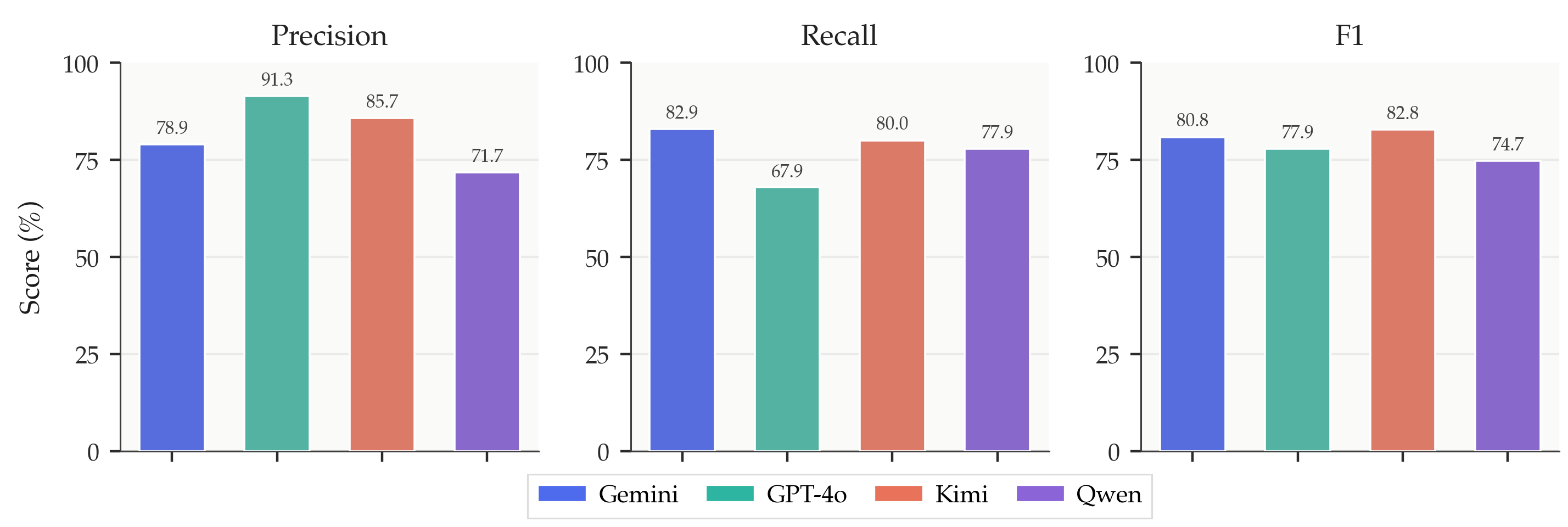}
    \caption{Precision, recall, and F1 scores for predicting primary motion graphics attributes across four LLMs: Gemini-3 Flash, GPT-4o, Kimi Code 2.5, and Qwen 3.}
    \label{fig:tech_results}
\end{figure}

\subsubsection{Method}
With ground truth established, we formulate attribute extraction as a structured JSON generation task and evaluate four LLMs (Gemini 3 Flash, GPT-4o, Kimi Code-2.5, and Qwen-3) on 50 p5.js scripts.  Our evaluation focuses on primary attributes, which define the core interaction experience, using micro-averaged Precision, Recall, and F1-score. Precision reflects the extent to which extracted attributes align with ground truth (avoiding unnecessary controls), Recall measures coverage of key parameters (avoiding missed functionality), and F1-score summarizes overall reliability in bootstrapping the on-demand UI.

\subsubsection{Attribute-Control Mapping Evaluation.}
\revise{Beyond attribute extraction, we further evaluated whether the system can map extracted attributes to appropriate controls. This evaluation differs from Fig.~\ref{fig:tech_results}, which measures whether an LLM can identify meaningful motion-graphics attributes. Here, we focus on the next step in the pipeline: selecting an appropriate UI primitive and interaction strategy for each attribute.}

\revise{Using the same 50-script benchmark, we sampled the expert-validated primary attributes and asked each model to generate an attribute--control mapping for every attribute. We compared two prompting conditions: (1) a direct prompting baseline, where the model selected controls from natural-language task descriptions alone, and (2) our skill-base guided condition, where the model was provided with structured examples and rules derived from commercial authoring tools. A mapping was counted as correct when the selected control and interaction strategy matched the expert-annotated control type or an equivalent authoring convention. For example, spatial position should be mapped to direct dragging or coordinate controls, color attributes to a color picker or palette, and temporal parameters to sliders or curve-based controls.}

\revise{Across LLMs, the skill-base guided condition substantially improved mapping precision, increasing the average precision from 53.5\% under direct prompting to 83.2\%. This result suggests that the skill base improves not only attribute coverage, but also the alignment between generated controls and established authoring practices.}

\end{document}